# Frontier vision–language models have overtaken young adults at detecting AI-generated portraits — but not their calibration

Sunwhi Kim[1†], Sunyul Kim[2], Meounggun Jo[3], Jini Tae[4]

[1] Hwasung Medi-Science University, Dept. of Bio-Healthcare, Republic of Korea

[2] Yonsei University, Graduate School of Engineering, Dept. of Artificial Intelligence, Republic of Korea

[3] Hoseo University, Republic of Korea

[4] Gwangju Institute of Science and Technology (GIST), School of Humanities and Social Sciences, Republic of Korea

*† Corresponding author*

## Abstract

AI image generators now create face portraits that are hard to tell from real photographs. Vision–language models (VLMs) are increasingly proposed to flag such images. We benchmarked 19 VLMs on the same 198 face portraits — real photographs and identity-matched ChatGPT-4o and Imagen 3 versions — under the same task as our earlier study of 1,667 adults (85% correct overall; accuracy fell steeply with age). The June-2026 cohort of 14 models only matched adults in their 20s–30s. Four weeks later the ceiling broke. Among five July-2026 releases under the identical protocol, gpt-5.6-sol reached 92.8% balanced accuracy (five-draw mean 92.1%), clearly above adults in their 20s (88.5%), and claude-fable-5 detected every AI image while averaging 91.9%. Model sensitivity now exceeds young adults decisively (d′ up to 3.4 versus ≈ 2.4). What has not been overtaken is human calibration. Model criteria spread from c = −1.10 to +1.45 while humans sit near zero at every age; both new leaders are biased (+0.44, −0.97), and only a few mid-ranked models approach the human balance. Changing the labelled examples still flipped about one answer in four. The best machines now out-see young adults here, without matching the human balance between suspicion and trust.

## Introduction

Generative models now produce face portraits with sufficient photorealism to circulate as profile pictures, advertisements, and news-like visuals [1,2], and AI-generated faces already populate fake social-media profiles at scale [3]. Human observers struggle with such content. Synthetic faces from the earlier GAN (generative adversarial network) era were judged indistinguishable from real photographs — and even more trustworthy [4–6]. Some AI faces are now perceived as human *more* often than actual human faces [7]. And across stimulus types, human detection of AI-generated media is modest, biased toward accepting content as authentic, and resistant to simple awareness interventions [8–12]. An earlier study of ours, so far reported only in a preprint [13], measured this ability for face portraits produced by generators available in July 2025. There, 1,667 adults aged 20–69 judged portraits one at a time, choosing "real" or "AI" on each trial (our analytic sample of that study's public data release; Methods). The portraits were real photographs from the public FFHQ (Flickr-Faces-HQ) dataset and identity-matched ChatGPT-4o and Imagen 3 versions of them. Mean accuracy was 85%. That is far above the near-chance pooled accuracy reported in a recent meta-analysis of human deepfake detection [11]. The gap plausibly reflects two factors. First, the task is narrow and clean. Each trial presented a single portrait for a binary REAL/AI judgment, drawn from a curated pool of real FFHQ photographs and identity-matched AI recreations of them at balanced base rates. This is not open-ended detection of arbitrary fakes in the wild. (Identity matching describes how the stimulus set was built; the versions of an identity were never shown together, and observers judged each image on its own.) The present benchmark supplies an internal check on this reading: judging the very same images, 19 frontier models spread from 58.3% to 92.8%, so the pool is not trivially easy. Second — a possibility our data cannot test — people may be adapting, learning the tells of synthetic media as it saturates everyday life. Crucially, the ability was far from uniform: accuracy fell from about 88% in the 20s to about 66% in the 60s. How well people detect AI faces therefore depends heavily on who is judging, which is central to the comparison we draw here.

A natural response to imperfect human detection is automation. Dedicated forensic classifiers detect generated images with high accuracy in-distribution, but their performance degrades on unseen generators and post-processing [14–17]. Large vision–language models (VLMs) offer a different promise: general-purpose visual reasoning that might transfer to forensic judgments without task-specific training. Recent evaluations have begun to probe multimodal models as zero- or few-shot detectors and explainers of synthetic imagery [18–22]. Implicit in much of this work — and in public expectation — is the assumption that frontier AI systems now exceed human perceptual abilities on such tasks.

Testing that assumption requires more than a model leaderboard. Model evaluations and human experiments typically use different stimuli, different response formats, and different scoring, so "model X beats humans" claims often rest on numbers measured under conditions that cannot be directly compared. The machine-behaviour tradition argues instead for evaluating AI systems with the same experimental instruments used for humans [23,24]. Careful psychophysics-style comparisons in vision have found that apparent model–human parity can dissolve once testing conditions are matched [25], often because models succeed through shortcuts humans do not use [26]. Language models also bring failure modes that human psychophysics does not anticipate. Their answers can change with superficial prompt variations, such as the choice or order of in-context examples [27–30]. Neural networks are systematically overconfident [31], and language models' verbalized confidence is often poorly calibrated even when their internal probabilities are informative [32–34]. And their stated reasons can misrepresent the actual basis of their answers [35,36], echoing a classic finding about human self-reports [37]. A meaningful human–model comparison should therefore measure not only accuracy but also robustness, bias, calibration, and the evidential status of model explanations.

Here we present such a comparison — twice. Nineteen frontier VLMs from nine providers judged 198 of the 210 portrait stimuli from our human study (all but the practice identities). A first cohort of 14 models was collected in June 2026. After a wave of releases about four weeks later, a second cohort of five July-2026 models (gpt-5.6-sol/terra/luna, kimi-k3, claude-fable-5) was collected under the identical protocol. The protocol mirrors the human task. Models saw the same four labelled practice images as few-shot examples (a handful of labelled cases shown before the test), made the same binary REAL/AI judgment on one image at a time, and were scored to reproduce the human session composition (10 real / 5 ChatGPT-4o / 5 Imagen 3). This enables a like-for-like comparison against the age-stratified human reference distribution. Beyond a single accuracy estimate, we measured (i) stability of each model's judgments across varied few-shot example sets, (ii) response bias, (iii) confidence calibration, (iv) the content and diagnostic value of 3,738 model-stated rationales, and (v) generator-specific difficulty.

The two cohorts tell a before-and-after story. In June 2026 no model surpassed the young-adult human ceiling: the best were statistically indistinguishable from adults in their 20s–30s. In July 2026 the ceiling broke — gpt-5.6-sol and claude-fable-5 exceeded the young-adult mean on single passes, on draw averages, and in sensitivity ($d'$). What did not change is the character of the machine failures: extreme response criteria, example-driven instability, and verdict-aligned rationales persisted in the new leaders. Re-running with different practice images changed roughly a quarter of per-item verdicts. Response biases reached extremes absent in humans, confidence was largely non-diagnostic, and the human-like cues models cited aligned with their verdicts rather than with the truth of the image.

## Results

### A human-comparable benchmark for vision–language models

Figure 1 summarizes the stimuli and the matched human–model protocol. The stimulus pool was identical to that of the earlier human study [13]: 210 face portraits (512×512 px). It comprised 70 real FFHQ photographs [1] and 140 AI counterparts, generated from the same identities in July 2025 by ChatGPT-4o (native image generation) [38] and Imagen 3 (via Gemini 2.5) [39] using a mirroring procedure (Methods). Each of the 19 VLMs (Table 1) was queried via OpenRouter at temperature 0, the setting that makes a model's answers most repeatable. Fourteen models from eight providers (OpenAI, Anthropic, Google, xAI, Alibaba, Zhipu, Mistral, Meta) were collected in June 2026, and five newly released models about four weeks later, one of them from a ninth provider (Moonshot). Each received the four labelled practice images from the human experiment as few-shot examples (2 real, 1 ChatGPT-4o, 1 Imagen 3), then judged the remaining 198 stimuli once — one image per query, exactly as human observers saw one image per trial. For each image the model returned a REAL/AI answer, a 0–100 confidence rating, and a one-sentence rationale. In the human study, each person completed a 20-trial session (10 real / 5 ChatGPT-4o / 5 Imagen 3). We scored each model the same way, as balanced accuracy — accuracy weighted to match that image mix (0.5·real + 0.25·ChatGPT-4o + 0.25·Imagen 3). To compare models and people, we then resampled many 20-trial sessions with the same mix (a method called bootstrapping; Methods).

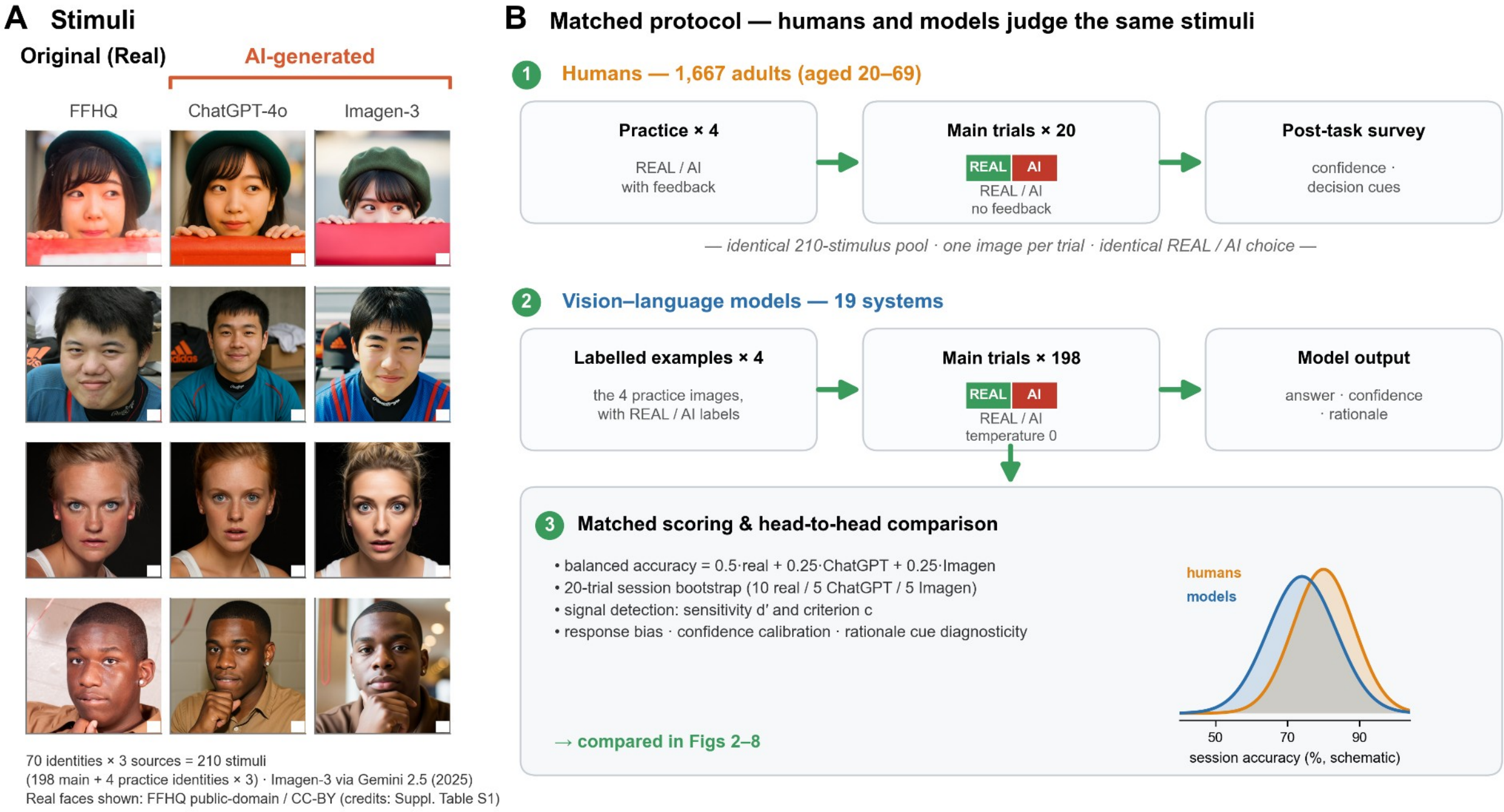

**Figure 1 | Stimuli and the matched human–model protocol.** (A) Stimulus generation, identical to the earlier human study [13]: 70 real FFHQ identities, each mirrored by ChatGPT-4o (native image generation) and Imagen 3 (via Gemini 2.5) in July 2025 (210 images; 198 main + 4 practice identities × 3 versions excluded). The rows shown here illustrate how the pool was built; during testing each image was judged on its own, and the three versions of an identity were never presented together. Example identities shown are limited to FFHQ images with public-domain/CC-BY licences (image-level credits in supplementary table S1 of ref. [13]). (B) The two observer types judge the same stimuli under matched protocols. Humans (lane 1): 1,667 adults completed 4 practice trials with feedback, then 20 main REAL/AI trials without feedback, then a survey (confidence, decision cues). Models (lane 2): each of 19 vision–language models (systems that read both images and text; 14 collected in June 2026, five July-2026 additions) first saw the same four labelled examples up front — a "few-shot prefix", i.e. a few solved cases shown before the test. Each then judged all 198 main stimuli once at temperature 0, returning a REAL/AI answer, confidence and one-sentence rationale. Both observers are scored the same way (3): balanced accuracy (accuracy weighted by image type; 0.5·real + 0.25·ChatGPT-4o + 0.25·Imagen 3), resampled 20-trial sessions (10 real / 5 ChatGPT-4o / 5 Imagen 3), and signal-detection sensitivity (d′) and bias (criterion c). These feed the head-to-head comparisons in figures 2–8. The session-accuracy densities are schematic.

**Table 1 | Model roster.** Provider, OpenRouter model identifier, tier, and list price per million tokens (input / output) at each cohort's collection time (June/July 2026).

| Model | Provider | OpenRouter identifier | Tier | In $/M | Out $/M |
|---|---|---|---|---|---|
| gpt-5.5 | OpenAI | openai/gpt-5.5 | flagship | 5.0 | 30.0 |
| gpt-5.4 | OpenAI | openai/gpt-5.4 | mid | 2.5 | 15.0 |

| Model | Provider | OpenRouter identifier | Tier | In $/M | Out $/M |
|---|---|---|---|---|---|
| gpt-5.4-mini | OpenAI | openai/gpt-5.4-mini | budget | 0.75 | 4.5 |
| claude-opus-4.8 | Anthropic | anthropic/claude-opus-4.8 | flagship | 5.0 | 25.0 |
| claude-sonnet-4.6 | Anthropic | anthropic/claude-sonnet-4.6 | mid | 3.0 | 15.0 |
| gemini-3.1-pro | Google | google/gemini-3.1-pro-preview | flagship | 2.0 | 12.0 |
| gemini-3.5-flash | Google | google/gemini-3.5-flash | fast | 1.5 | 9.0 |
| gemma-4-31b | Google | google/gemma-4-31b-it | open | 0.12 | 0.35 |
| grok-4.3 | xAI | x-ai/grok-4.3 | flagship | 1.25 | 2.5 |
| qwen3-vl-235b | Alibaba | qwen/qwen3-vl-235b-a22b-instruct | open-VL | 0.2 | 0.88 |
| qwen3.5-flash | Alibaba | qwen/qwen3.5-flash-02-23 | fast | 0.07 | 0.26 |
| glm-4.6v | Zhipu | z-ai/glm-4.6v | open-VL | 0.3 | 0.9 |
| mistral-medium-3.1 | Mistral | mistralai/mistral-medium-3.1 | mid | 0.4 | 2.0 |
| llama-4-maverick | Meta | meta-llama/llama-4-maverick | open | 0.15 | 0.6 |
| gpt-5.6-sol | OpenAI | openai/gpt-5.6-sol | flagship | 5.0 | 30.0 |
| gpt-5.6-terra | OpenAI | openai/gpt-5.6-terra | mid | 2.5 | 15.0 |
| gpt-5.6-luna | OpenAI | openai/gpt-5.6-luna | budget | 1.0 | 6.0 |
| kimi-k3 | Moonshot | moonshotai/kimi-k3 | flagship | 3.0 | 15.0 |
| claude-fable-5 | Anthropic | anthropic/claude-fable-5 | flagship | 10.0 | 50.0 |

## The young-adult ceiling held in June 2026 — and broke in July

Ranking the 19 models together with the five human age groups produces the leaderboard in Fig. 2 (the human age gradient alone is shown in Supplementary Fig. S1). The June-2026 cohort stopped at the human ceiling. Its best models — gpt-5.5 (87.1%; 95% confidence interval ±4.2 percentage points, pp), qwen3-vl-235b (87.1%) and claude-sonnet-4.6 (86.7%) — reached the human average (85.18%) but were statistically indistinguishable from the 20s–30s means (88.5%/87.7%). An identity-cluster check gave the same margins (e.g. gpt-5.5 between 83.0% and 91.3%; Methods). The July-2026 cohort broke it. gpt-5.6-sol scored 92.8% (cluster CI 89.8–95.8%), 4.3 pp above the 20s mean; at maximum reasoning effort it reached 94.3% with zero false alarms (pilot; Methods). claude-fable-5 scored 88.6% while detecting every AI image it saw (132/132); its cost was real-photo accuracy, 77.3%. Pairwise McNemar tests now separate the new leaders from the old ones: gpt-5.6-sol beat gpt-5.5 ($p = .011$) and qwen3-vl-235b ($p = .019$), and claude-fable-5 beat both as well ($p = .023, .026$); sol and fable-5 could not be separated from each other (Supplementary Fig. S2). Humans in their 20s (88.5%) now rank third. Adults in their 40s (80.5%) fell below most of the leading models, and every model in the top half outperformed the 50s (71.8%) and 60s (65.9%). At the other end, the four lowest-scoring models trailed even the 60s mean (65.9%), down to 58.3% for llama-4-maverick. kimi-k3, reportedly dominant on coding benchmarks, placed 12th of 19 at 80.9% (with 21 responses unparseable at the token cap). Model scale, price, and recency still did not guarantee performance: the flagship claude-opus-4.8 (64.0%) trailed its mid-tier sibling by 22.7 pp, and gpt-5.6-terra (87.1%) merely tied the June leaders.

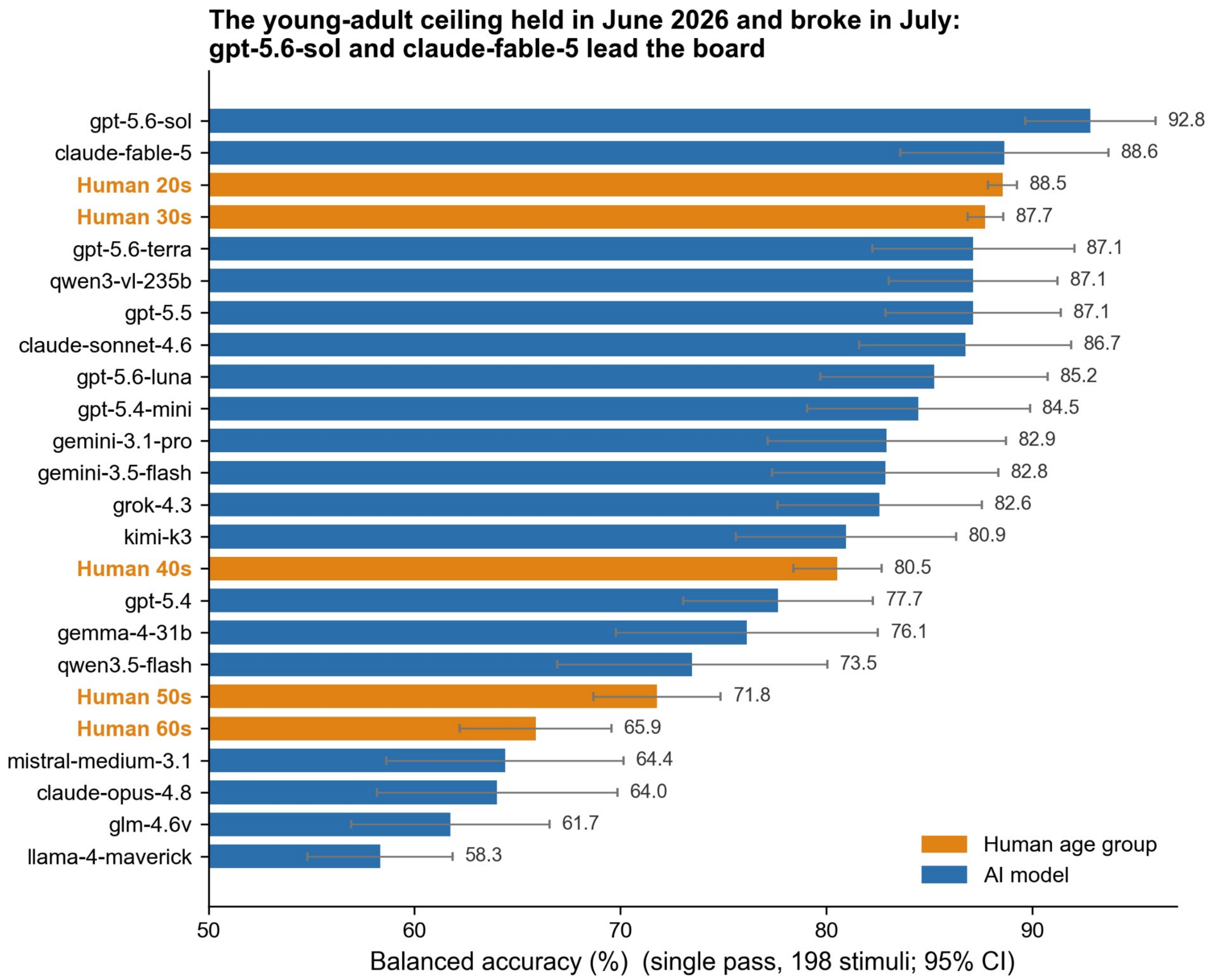

**Figure 2 | The young-adult ceiling held in June 2026 and broke in July.** Single-pass balanced accuracy (0.5·real + 0.25·ChatGPT-4o + 0.25·Imagen 3) for 19 models, shown alongside the five human age groups (orange). Error bars are 95% confidence intervals — the range in which the true score likely lies (models: delta-method propagation of the per-class binomial intervals; humans: ±1.96 standard errors of the mean, SEM). gpt-5.6-sol tops the board with its interval lower bound above the human 20s mean; claude-fable-5 matches the 20s single-pass; the June-2026 leaders' intervals overlap the 20s–30s means (no clear separation), and the leading models' interval lower bounds sit above the means of the groups aged 40 and over. Human data reused from ref. [13].

## Instability persists: few-shot examples still destabilize model judgments

A single deterministic pass conflates a model's ability with its luck. At temperature 0, the only element of the protocol we vary is which four practice images precede the trials. We therefore repeated the full benchmark five times per model, each time drawing a different labelled practice set (2 real, 1 ChatGPT-4o, 1 Imagen 3, rotated out of the 70-identity pool) while keeping instructions fixed (Methods).

Averaging over draws splits the cohorts again (Fig. 3a). No June-2026 model reached the human mean this way: that cohort's top cluster converged to 83.6–84.5% (gpt-5.4-mini 84.5 ± 3.1; gpt-5.5 84.4 ± 6.1; claude-sonnet-4.6 84.1 ± 3.5; qwen3-vl-235b 83.6 ± 2.4; mean ± standard deviation, SD). The July leaders held above the ceiling: gpt-5.6-sol averaged 92.1 ± 4.0 and claude-fable-5 91.9 ± 2.3. For fable-5, the hand-selected baseline draw was in fact its worst pass. The crossing is therefore not a single-pass accident. Instability itself, however, did not disappear. gpt-5.5 ranged from 72.3% to 89.0% across draws, single-pass scores within one model spanned up to 18.2 pp (gpt-5.4: 64.0–82.2%), and even gpt-5.6-sol dropped to 84.5% — the June-cohort level — on one draw. (Each draw changes both the example set and, slightly, the 198 judged items, so these spreads are an upper bound on the example effect; Methods.)

We also measured the flip rate: the share of items whose answer changed at least once across the six passes — the baseline plus the five varied draws (Fig. 3b). Across the 19-model fleet this was 25.7% — roughly one image in four (strict REAL↔AI reversals: 24.7%; strict reversals across the five varied draws alone: 23.3%). How much of this is the examples, and how much the models' own run-to-run noise? An identical-prompt control estimated the two components (Methods). For the stable models, the like-for-like two-pass example effect was several times their noise (claude-sonnet-4.6: 6.8% vs 1.5%; qwen3-vl-235b: 6.9% vs 2.0%). For the two June OpenAI models tested, the example effect barely exceeded the noise (gpt-5.5: 11.4% vs 9.1%; gpt-5.4: 14.4% vs 11.6%); much of their volatility is intrinsic run-to-run noise rather than example sensitivity. Sensitivity clustered by family. OpenAI's largest June models had the widest accuracy spreads (gpt-5.4: SD ±6.8, 34.8% flips; gpt-5.5: ±6.1, 31.0%), and gpt-5.6-terra continued the pattern (36.7% flips). The Claude family was the most stable — claude-fable-5 set the fleet record at 11.4% flips, with the June Claude models next (14.8% each). gemma-4-31b and qwen3.5-flash combined low accuracy with extremely stable overall scores (SD ±0.9). Yet their per-item answers flipped at above-average rates (29% and 32%); the flips largely cancelled out in the score. Accuracy and robustness are therefore separable properties, and single-pass leaderboards overstate both the level and the reliability of model performance.

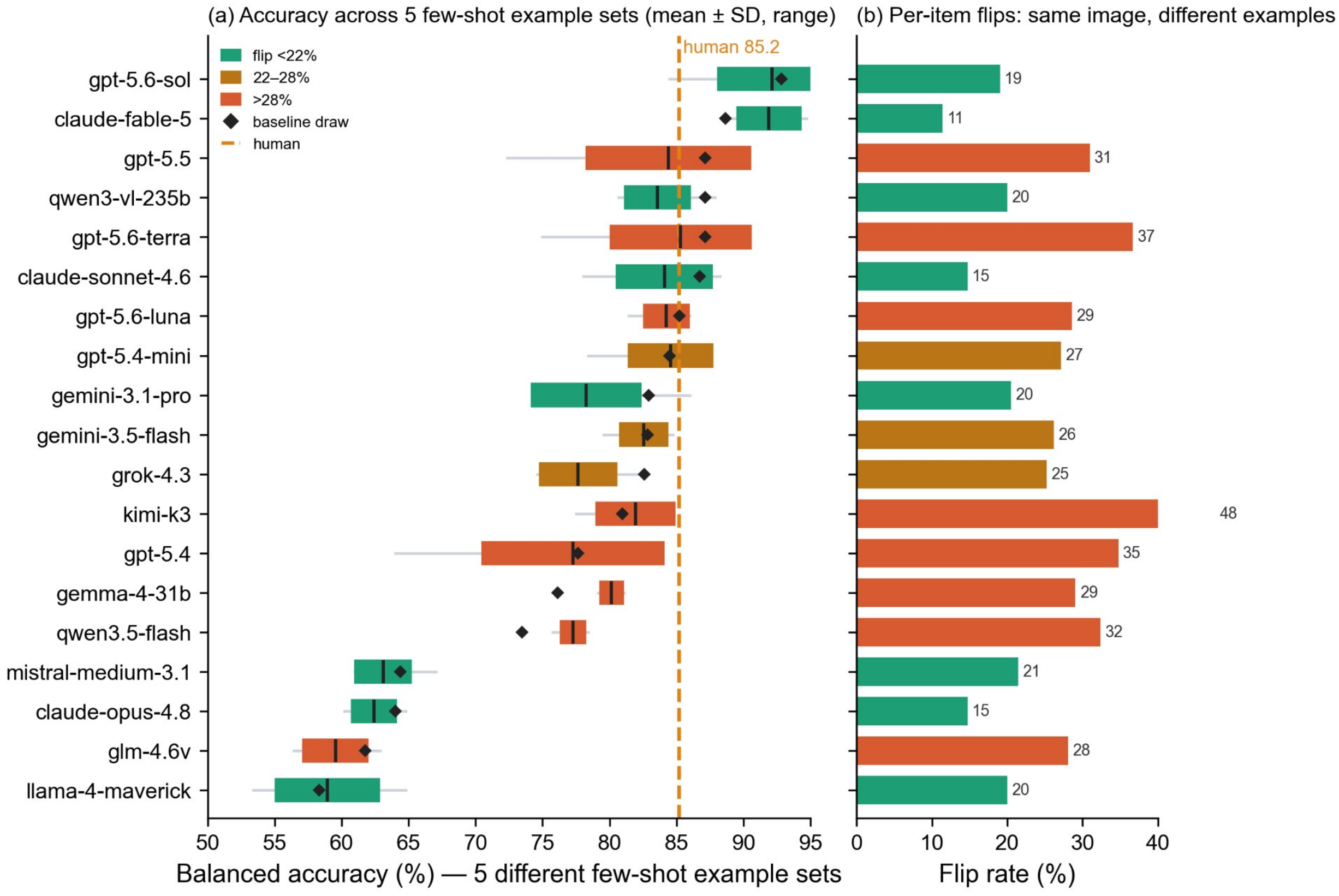

**Figure 3 | Model judgments are fragile to the few-shot examples.** (a) Balanced accuracy per model across 5 varied labelled-example draws (box: mean ± SD; whisker: range; ◆: the fixed baseline draw; dashed line: human mean 85.18%; box colour: flip rate). The spread mixes the example effect with the slightly different 198-item set judged per draw and is an upper bound on example-sensitivity. (b) Per-item flip rate — the fraction of items whose answer changed at least once across the six passes (the baseline plus the five varied draws); fleet mean 25.7% over the 19 models. This includes the models' run-to-run noise as well as the example effect (Methods).

## Models exhibit extreme, non-human response biases

The human observers were nearly unbiased, with real-photo accuracy exceeding AI-detection accuracy by only 1.0 pp (85.7% vs 84.7%). Models spanned a bias spectrum from +64 to −80 pp (Fig. 4a). claude-opus-4.8 called almost everything AI (real-photo accuracy 31.8%, AI detection 96.2%; bias +64.4 pp; 86.9% of all answers were "AI"). llama-4-maverick showed the mirror-image failure, essentially rubber-stamping images as real (real 98.5%, ChatGPT-4o detection 9.1%; bias −80.3 pp). glm-4.6v (−64.4), mistral-medium-3.1 (−47.0) and gpt-5.4 (−38.6) also leaned heavily toward "REAL". Only a minority of models approximated the human balance (gemini-3.5-flash −0.9 pp; gpt-5.6-terra −1.5 pp; gpt-5.4-mini +2.3 pp; claude-sonnet-4.6 +6.8 pp). The new leaders were not among them: claude-fable-5 bought its perfect AI detection with a 22.7-pp bias toward "AI" (it called 22.7% of real photographs AI), and gpt-5.6-sol leaned 11.4 pp the other way. Raw accuracy conceals these failures — a strongly biased model can still score well on one class — which is why all headline accuracy rankings use class-balanced accuracy.

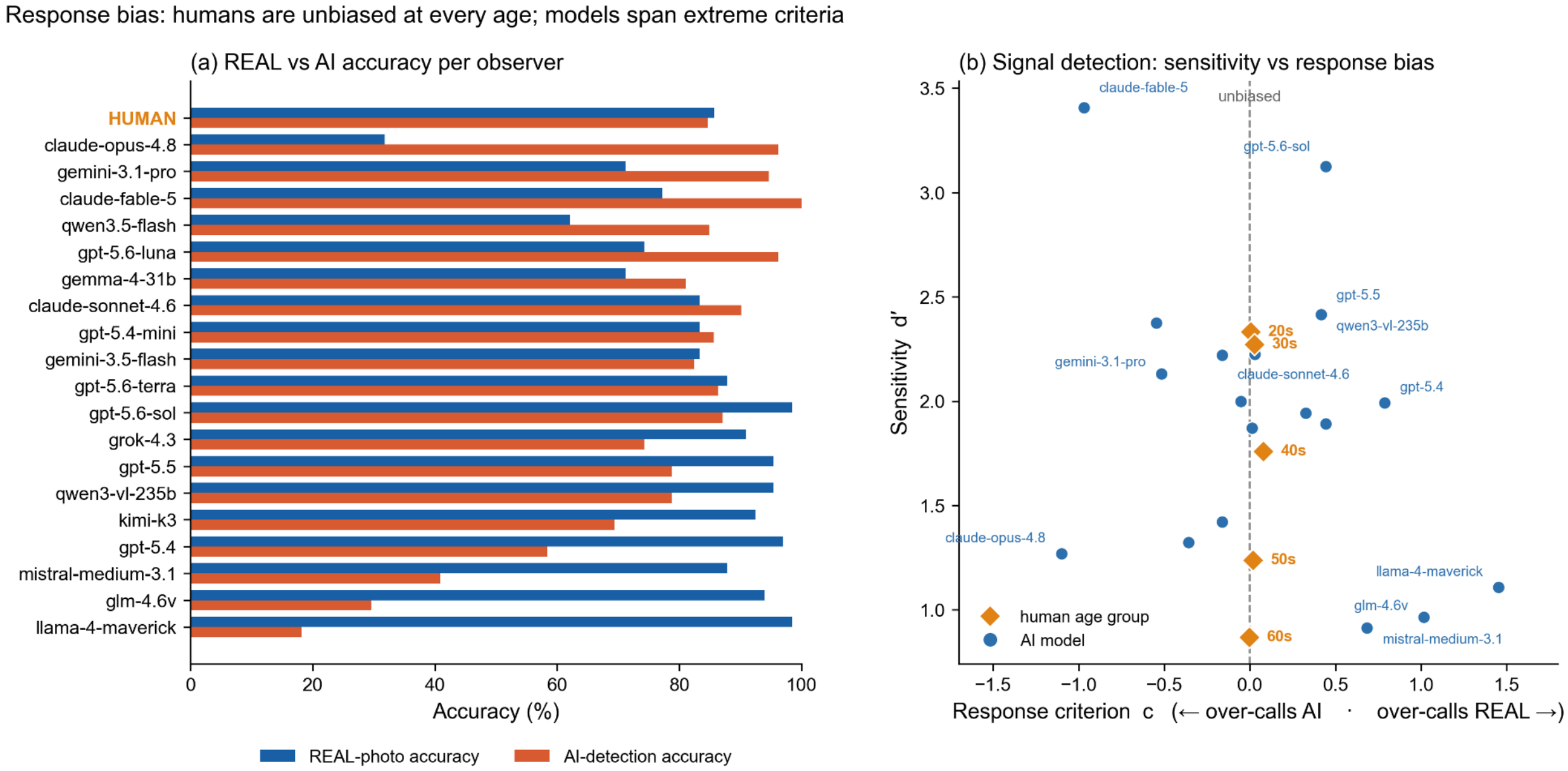

**Figure 4 | Response bias and signal detection.** (a) Real-photo accuracy versus AI-detection accuracy per model, with the human reference in the top row. Model biases span +64.4 pp (claude-opus-4.8, over-calls AI) to −80.3 pp (llama-4-maverick, over-calls REAL); humans are nearly balanced (−1.0 pp). (b) Sensitivity d′ versus response criterion c, with AI as the target signal. Here d′ is how well an observer separates real from AI, and c is which answer they lean toward ($c \approx 0$ = even-handed). Human age groups (orange diamonds) form a vertical column at $c \approx 0$, with d′ falling from the 20s to the 60s. Models (blue) spread across the bias scale ($c = -1.10$ to $+1.45$). The July leaders exceed young-adult sensitivity (d′ up to 3.4) but carry response biases humans lack. Human points are participant-mean d′/c; each model is a single-observer d′/c from its pooled counts — both log-linear corrected. Human points recomputed from the public data release of ref. [13] (Methods).

A signal-detection analysis (Fig. 4b) separates two things: sensitivity (d′), how well an observer tells real from AI, and bias (criterion c), which answer the observer leans toward. Treating AI as the signal to detect, the human age groups sit in a near-vertical column at an unbiased criterion ($c \approx 0$ from the 20s to the 60s), while their sensitivity d′ falls with age. (We recomputed the human d′ and c here from the public data release of ref. [13].) Models, by contrast, spread along the bias scale, from claude-opus-4.8 at $c = -1.10$ to llama-4-maverick at $c = +1.45$. The new leaders are not at the neutral point either: claude-fable-5 sits at $c = -0.97$, nearly matching the most AI-leaning model in the fleet, while gpt-5.6-sol leans the other way at $c = +0.44$. Humans never leave $c \approx 0$. Sensitivity is where the July cohort changed the story. The June leaders were statistically at the young-adult level (gpt-5.5 and qwen3-vl-235b d′ = 2.42 corrected, cluster CI [2.0, 3.2]; young adults: participant-mean 2.33, pooled-count 2.41). The new leaders exceed it decisively: gpt-5.6-sol d′ = 3.13 (corrected; CI [2.65, 3.75]) and claude-fable-5 d′ = 3.41 (CI [3.10, 3.75]) — both interval lower bounds clear the young-adult value. Sensitivity is no longer the frontier's limit; the criterion is. fable-5's $c = -0.97$ turns superhuman discrimination into a balanced score (88.6%) no better than a human twenty-something: the balanced task gives real and AI images equal total weight, so a biased model wastes what it sees. gpt-5.6-sol wins precisely because its bias, while present ($c = +0.44$), is mild enough not to squander what it sees. In short: the leading machines now out-see young adults, and what still separates them is where they place the decision criterion.

## Model confidence: inflated in the June cohort, improving in the July leaders

Humans' self-rated detection confidence tracked their actual accuracy (Spearman $\rho = 0.379$, $n = 1,667$; recomputed here from the public data of ref. [13]). That is a between-participant correlation, whereas the model measures below are within-model, per-judgment quantities; the contrast is therefore qualitative. Model confidence behaved differently on both axes we measured (Fig. 5), though here the July cohort shows real movement. First, calibration. Most June models were overconfident, with mean stated confidence exceeding balanced accuracy by up to +29 pp (llama-4-maverick: mean confidence 87.5 vs 58.3% accuracy), +28 pp (mistral-medium-3.1) and +22 pp (glm-4.6v). Only gpt-5.5 (−6.8), grok-4.3 (−5.1) and claude-sonnet-4.6 (−3.3) were not. In the July cohort, calibration improved markedly: gpt-5.6-sol's gap was +0.1 pp, and claude-fable-5 (−5.7), gpt-5.6-terra (−5.1) and kimi-k3 (−0.8) sat at or below zero (gpt-5.6-luna: +3.3). Second, resolution — whether higher confidence actually marks a more likely-correct answer. In the June cohort this was largely absent (differences between −4.1 and +7.8 pp; most within ±2 pp). The July models delivered the largest resolutions we observed: gpt-5.6-sol +8.5, claude-fable-5 +9.3 and gpt-5.6-luna +12.4 pp, against a June best of +7.8 pp (claude-sonnet-4.6). The Brier score (lower is better) summarized the shift: the two July leaders set new bests (claude-fable-5 0.070, gpt-5.6-sol 0.073, versus the June best of 0.098), while llama-4-maverick stayed worst (0.437). Metacognition, unlike the decision criterion, is improving with the frontier. High stated confidence still accompanied errors, though: in the fooled example of Fig. 6a, a June top-3 model accepted an AI-generated portrait as natural at confidence 95.

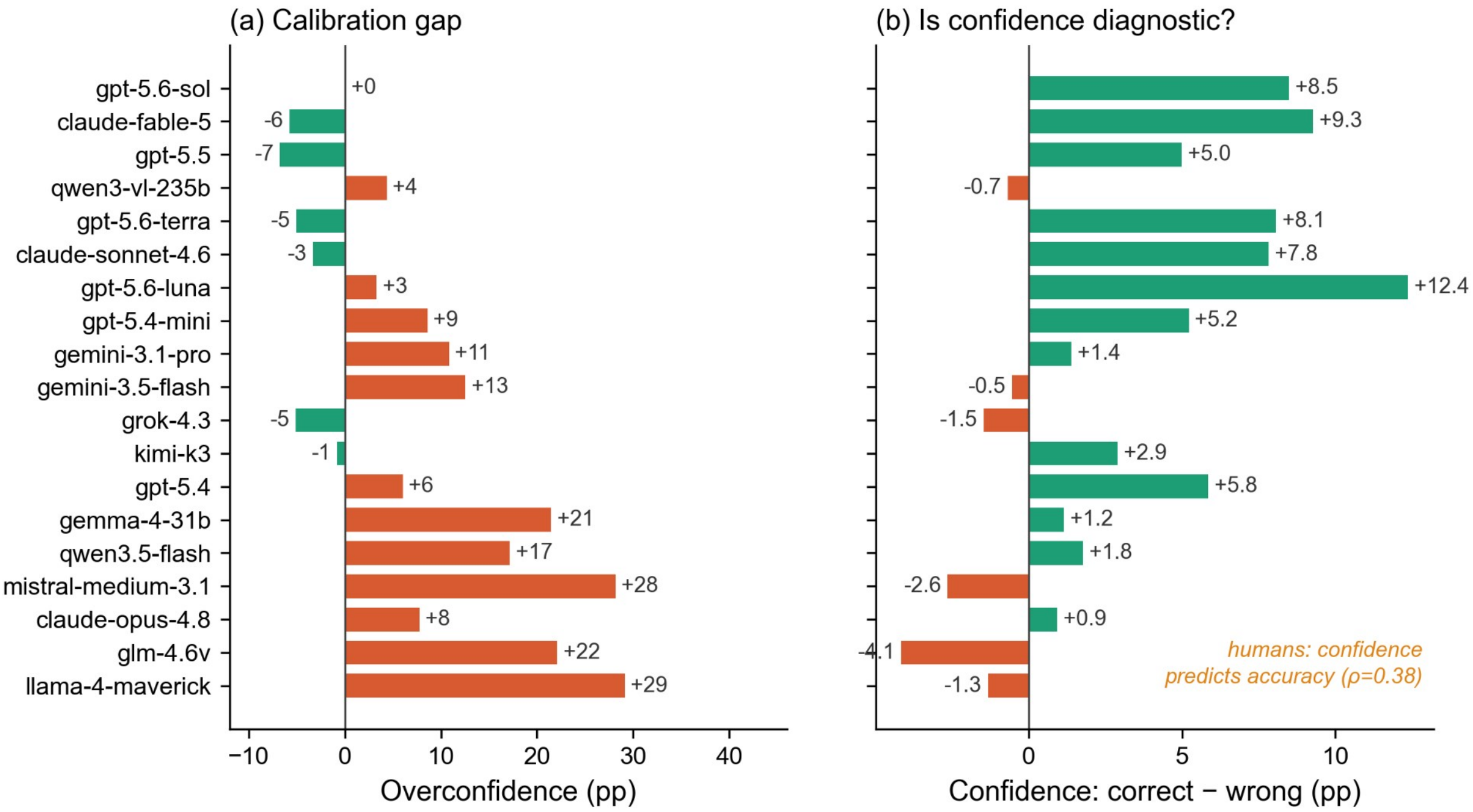

**Figure 5 | Metacognition.** (a) Overconfidence — mean stated confidence minus balanced accuracy, i.e. how much a model's confidence exceeds how often it is right. Most June-2026 models were overconfident, by up to +29 pp; the July-2026 leaders were nearly calibrated (gpt-5.6-sol +0.1 pp). (b) Resolution — mean confidence on correct answers minus that on wrong ones, i.e. whether a model is surer when it is right. Near zero for most June models (June best +7.8 pp); the July models show the largest resolutions observed here (up to +12.4 pp). Humans: confidence tracked accuracy (between-participant correlation $\rho$ = 0.379, n = 1,667; recomputed from the public data of ref. [13]).

## Models cite human-like cues…

The 19 models produced 3,738 free-text rationales, which we coded with a 15-category cue taxonomy using a neutral large language model (LLM) coder (Methods). The resulting cue landscape (Fig. 6b) overlaps substantially with the strategies humans self-reported in the reference study (Fig. 6c). The two sides are counted differently — model values are shares of rationales, human values are shares of participants ticking a checklist item — so the comparison is qualitative. Models most often cited over-smoothness (49.4% of rationales), skin texture (45.4%), an uncanny/too-perfect appearance (45.3%), lighting and shadows (43.0%), and a natural/realistic overall impression (42.9%). These echo the texture (65.3% of participants) and painting-like appearance (68.6%) cues that dominated human reports. Of the nine visual cues on the human checklist, however, texture was the only one positively associated with human accuracy once age, sex and device were controlled (recomputed here from the public data, excluding responses with the checklist fault described in Methods; n = 1,465: +3.8 pp, odds ratio 1.36, 95% CI [1.23, 1.50]). Endorsing a painting-like appearance was not (OR 1.11, 95% CI [0.99, 1.23]); two cues (eyes, abnormal text) were negatively associated, and age was by far the strongest predictor. But the overlap is selective. Models mentioned eyes in only 10.3% of rationales, whereas 42.8% of human participants reported using them. And while 49.1% of humans endorsed feeling/intuition — an explicit admission of uncertainty — expressions of intuition or uncertainty were essentially absent from model rationales (0%). Every model verbalized its guesses as confident perceptual analysis.

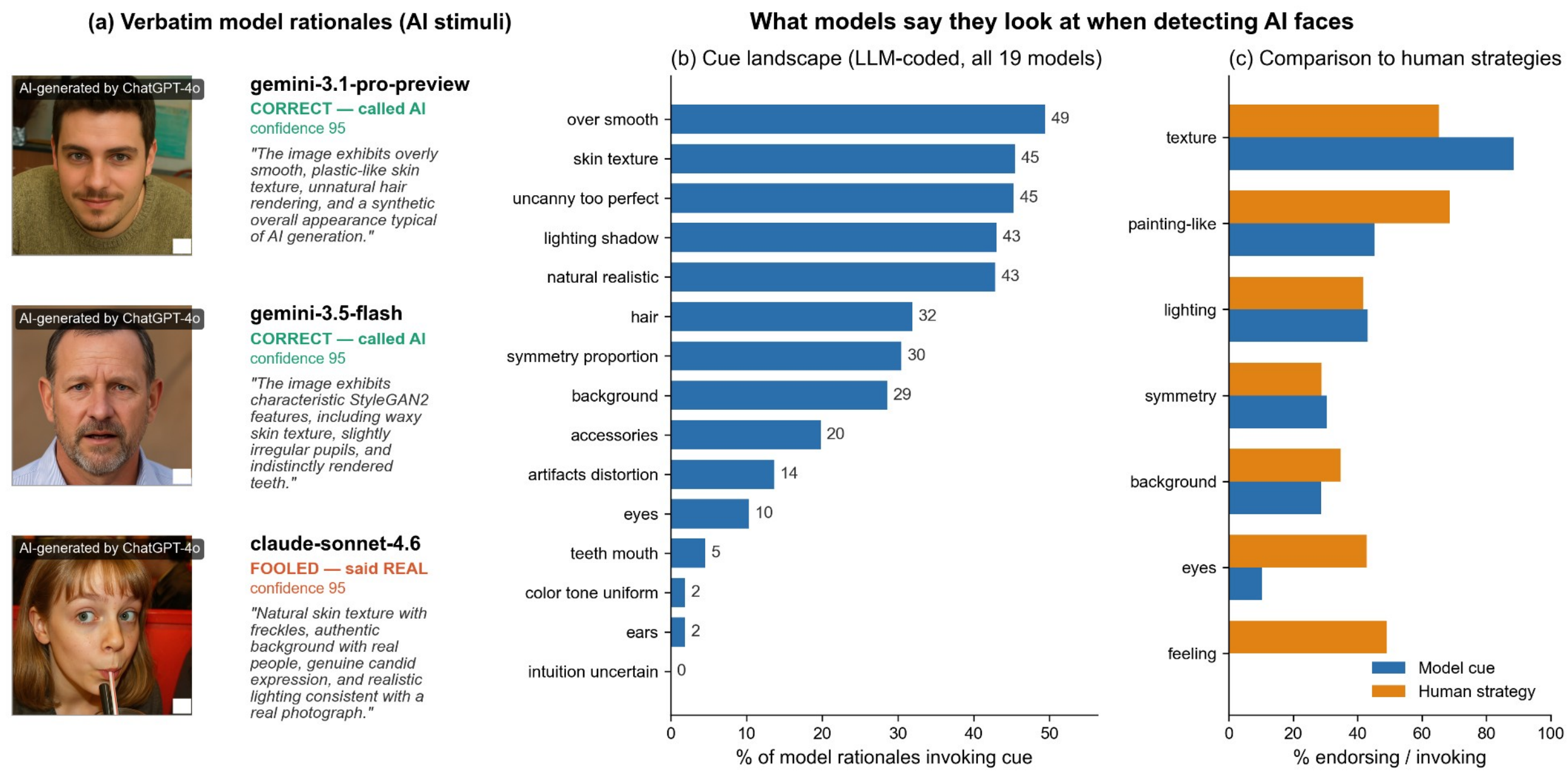

**Figure 6 | What models say they look at.** (a) Verbatim example rationales on AI-generated stimuli (stimulus shown with generator tag and model confidence): two models correctly flag AI images via concrete cues — one while misattributing the generator (StyleGAN2), itself a hint that stated cues are unreliable — and a top-3 model is confidently fooled (confidence 95), describing an AI portrait as natural/authentic. (b) Cue landscape of this study: frequency of 15 LLM-coded cue categories across all 3,738 model rationales. (c) Comparison to human self-reported strategies (ref. [13]), sorted by model usage: models over-cite smoothness/texture, under-cite eyes (10.3% vs 42.8%), and never express uncertainty (0% vs 49.1%). Model values are shares of rationales; human values are shares of participants (checklist) — a qualitative comparison.

## …but the cues are verdict markers, not evidence

Cue–verdict coupling is expected for any observer describing a decision it has just made. Two questions are informative: whether a cited cue predicts accuracy regardless of the image's true class, and whether it carries information beyond the verdict itself. On the first question, a cue's usefulness — how often citing it went with a correct answer — flipped depending on the image's true type, real or AI (Fig. 7). Overall, "reliable tells" such as uncanny/too-perfect (+13.4 pp accuracy when invoked vs base rate) and over-smoothness (+13.2) contrast with "fooled words" such as natural/realistic (−14.4). Conditioning on the true class exposes the mechanism. Over-smoothness was associated with +24.4 pp accuracy on AI images but −77.6 pp on real photographs. Natural/realistic showed the mirror pattern (−73.0 pp on AI images, +17.1 pp on real photographs). These cues track the verdict rather than the truth: a model that answers "AI" describes the image as too smooth, and one that answers "REAL" describes it as natural. Across the taxonomy, only hair-related cues retained a positive association in both classes (+2.7 pp within AI, +1.3 pp within real). On the second question, conditioning on the verdict itself revealed a gradient rather than uniform rationalization. Among REAL verdicts, the point estimate for citing natural/realistic was small (+3.5 pp relative to REAL verdicts without it), though this comparison is imprecise — only 62 REAL verdicts lacked the cue (95% CI roughly ±13 pp). Concrete surface cues, by contrast, retained modest conditional validity: among AI verdicts, over-smoothness +12.2 pp and uncanny/too-perfect +9.8 pp; among REAL verdicts, skin texture +25.9 pp (Methods). The dominant impression-level vocabulary of model rationales is therefore verdict-aligned commentary rather than demonstrable evidence: a single generated output cannot show whether the cue produced the verdict or merely accompanies it. This is consistent with reports that language-model explanations can systematically misrepresent the basis of the model's answers [35,36].

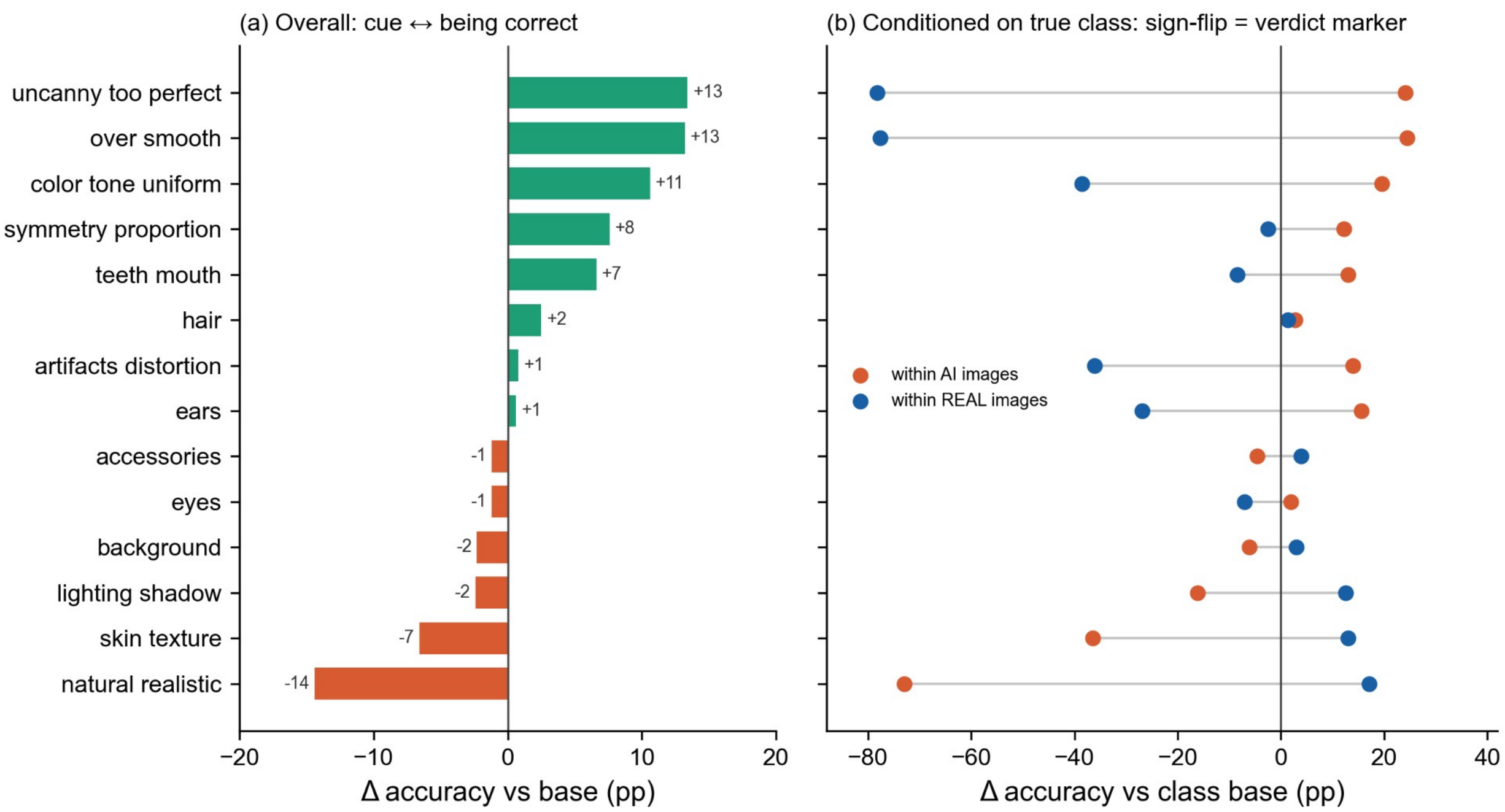

**Figure 7 | Rationale cues are verdict markers.** (a) Δ accuracy when a cue is invoked versus base rate, overall. (b) The same Δ, now split by the image's true class (real or AI). The main cues flip sign between classes (over-smoothness +24.4 pp within AI, −77.6 pp within real; natural/realistic −73.0 within AI, +17.1 within real). This means a cue mostly labels the model's verdict rather than providing evidence. Only hair is (weakly) positive in both classes.

## Models inherit the human difficulty ordering across generators

For the human observers, ChatGPT-4o portraits were substantially harder than Imagen 3 portraits (80.4% vs 88.9%). Fifteen of the 19 models reproduced this ordering (Fig. 8), often by wide margins (e.g., gpt-5.5: 71.2% vs 86.4%; grok-4.3: 60.6% vs 87.9%; gpt-5.6-sol: 80.3% vs 93.9%; kimi-k3: 67.2% vs 71.7%). claude-fable-5 tied the two generators at 100% each. The three exceptions were claude-opus-4.8 (97.0% vs 95.5%; consistent with its extreme AI bias) and two near-parity models, gpt-5.4-mini (86.4% vs 84.8%) and mistral-medium-3.1 (42.4% vs 39.4%). The relative detectability of generator families is thus largely a property of the stimuli that transfers across observer types, human or artificial.

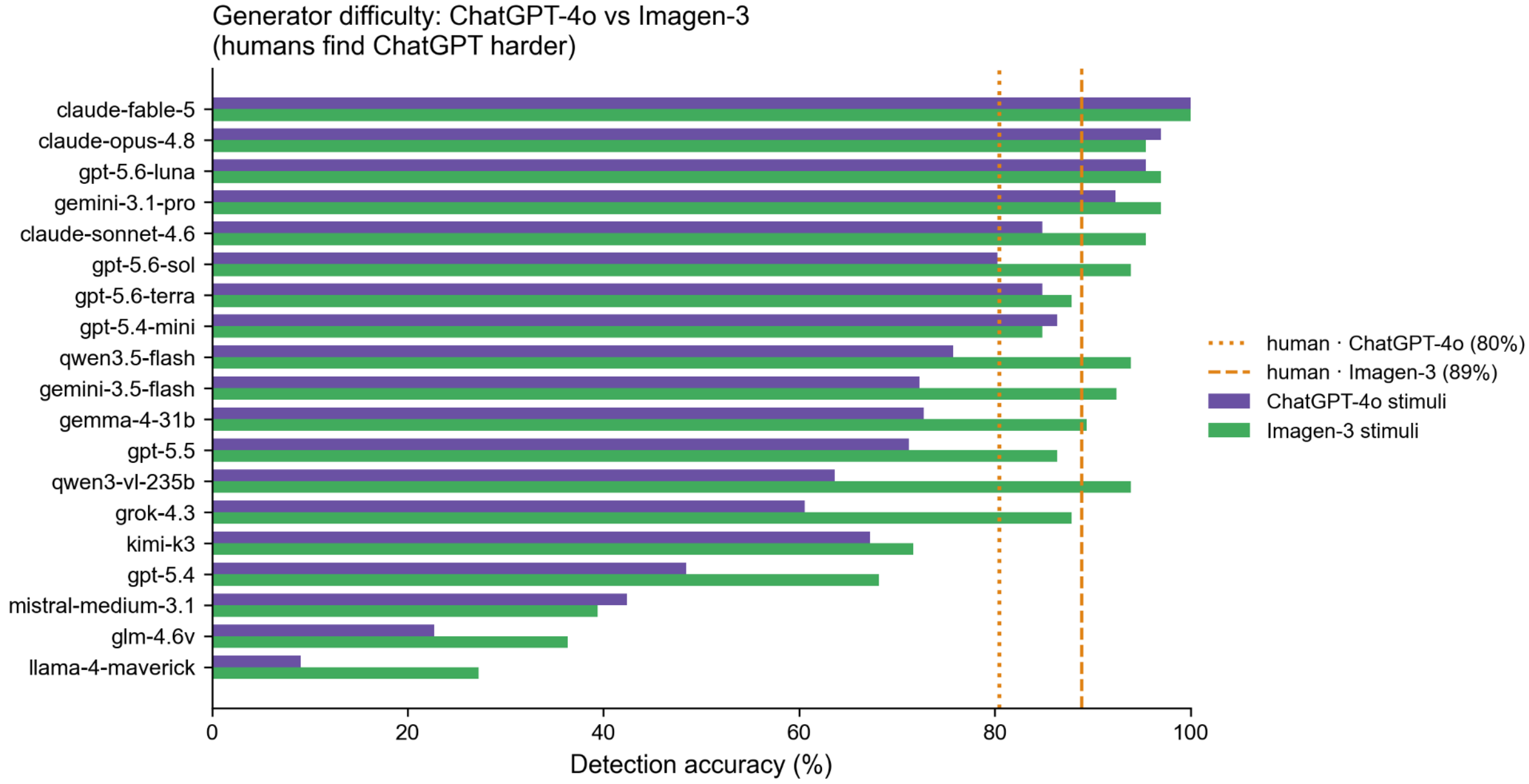

**Figure 8 | Generator difficulty transfers from humans to models.** Detection accuracy on ChatGPT-4o images versus Imagen 3 images, per model. 15 of 19 models find ChatGPT-4o images harder to spot, as humans did (80.4% vs 88.9%); claude-fable-5 ties the two at 100%. The exceptions are claude-opus-4.8, gpt-5.4-mini and mistral-medium-3.1.

### Age-persona prompting does not reproduce the human age gradient

Because the human reference is age-stratified, one may ask whether prompting a model to answer "as a typical human participant in their 20s/60s" would reproduce the human age gradient. We ran a screening pilot with pre-specified stop criteria: four models spanning the accuracy and bias spectrum (qwen3-vl-235b, gpt-5.5, claude-sonnet-4.6, llama-4-maverick), two persona conditions, and a 60-stimulus subset (Methods). Persona prompts produced only small and inconsistent shifts. All balanced-accuracy changes versus the neutral prompt were ≤3.8 pp, said-AI-rate changes ≤3.3 pp, and per-item flips ≤6.7%. gpt-5.5 even shifted in the direction opposite to human ageing (90.0% under the 60s persona vs 87.5% under the 20s persona; Supplementary Fig. S3). Demographic role-play therefore did not induce a human-like age gradient. We treat the age-stratified human data as a reference distribution, not a target for model role-play. Larger persona effects, if they existed, would in any case conflate perceptual change with prompt compliance and stereotype activation.

## Discussion

Frontier vision–language models have overtaken young adults at detecting AI-generated portraits — and our two-cohort design dates the crossing to within four weeks. Against an age-stratified human baseline collected under the same protocol and stimuli, the best of 14 June-2026 models were statistically indistinguishable from adults in their 20s–30s, and none exceeded the all-ages human mean when averaged over example draws. Five July-2026 releases later, gpt-5.6-sol (92.8% single-pass, 92.1% draw-averaged) and claude-fable-5 (91.9% draw-averaged, with perfect AI detection) sit clearly above the 20s mean, and both separate from the June leaders in paired tests. On this stimulus set, the best detector of AI portraits is, for the first time, a machine rather than the average adult of any age we measured. What our data equally show is what has not been overtaken. Neither new leader places its decision criterion where the human age groups place it (across the fleet, only a handful of mid-ranked models do); the example-driven instability that plagued the June cohort persists; and model rationales still track verdicts rather than truth. The gap between machines and people on this task has moved from *whether they can see* to *how they decide*.

Three characteristics keep the machine victory qualified. First, it remains *unstable*. Re-running the benchmark with different labelled example images changed the verdict on about a quarter of the images fleet-wide (a figure that includes the models' own run-to-run noise; Methods). It moved one June model's single-pass score across an 18-pp range, and dropped even gpt-5.6-sol to June-cohort level on one draw. This extends prompt-sensitivity findings from text benchmarks [27–29] to a perceptual task with near-deterministic decoding. Single-pass leaderboards — the dominant evaluation practice — can therefore overstate not just how well models perform but how dependably they do so. Stability and accuracy were also separate traits. The Claude family was the most stable, with claude-fable-5 setting the fleet record (11.4% flips). OpenAI's flagships were the least stable at similar accuracy; for the two June flagships covered by the control, much of that instability was intrinsic run-to-run noise rather than example sensitivity (Methods). And two low-accuracy models kept nearly constant overall scores even though their individual answers flipped often — the flips cancelled out. Reporting distributions over prompt draws, rather than point estimates, seems a minimal requirement for claims about model perceptual ability.

Second, the failure modes remain *non-human*. Signal detection makes the difference precise. Humans held an unbiased criterion ($c \approx 0$) at every age and simply lost sensitivity ($d'$) as they aged (recomputed from the public data of ref. [13]; Methods). Models — now including two whose sensitivity clearly exceeds young adults — carry response criteria spanning $c = -1.10$ to $+1.45$. The starkest case is the new one: claude-fable-5 sees better than the average adult of any age group we measured ($d' = 3.41$ versus 2.33 for the 20s; $c = -0.97$) yet converts that into a merely human-level balanced score, because it calls almost a quarter of real photographs AI. Bias across the fleet covered a 145-pp-wide range (+64 to −80 pp); one flagship answered "AI" for 87% of images while an open model accepted nearly every image as real. Confidence, unlike the criterion, is improving. June-cohort confidence was inflated by up to +29 pp and barely separated correct from wrong answers, echoing reports that language models' stated confidence is often poorly matched to their accuracy [33,34]. The July leaders, by contrast, were nearly calibrated and showed the largest resolutions we observed (human comparison: between-participant $\rho$ = 0.379; a qualitative contrast). In real use, a strongly biased detector still fails unevenly — flagging real photographs as fake, or letting fakes pass — in ways human raters would not, however well calibrated its confidence reports have become.

Third, model *explanations track verdicts, not evidence*. The cue vocabulary models used overlaps with the strategies human observers reported — texture, over-smoothness, lighting, painting-like impressions — which superficially suggests convergent perceptual strategies. But the usefulness of the main cues reversed with the true image. Take "over-smooth skin": models cite it both when they correctly flag AI images and when they badly misjudge real photographs. These cues are therefore verdict markers — language aligned with the answer rather than demonstrable evidence for it. This aligns with earlier work: model-stated reasoning can fail to reflect the model's real basis for an answer [35,36,40], and people's spoken explanations often misdescribe why they judged as they did [37]. The same caution applies to the human side, and our recomputation makes its limits concrete. Of the nine visual cues on the human checklist, texture was the only positive predictor of accuracy once age, sex and device were controlled (odds ratio 1.36, +3.8 pp); two cues (eyes, abnormal text) predicted it negatively, and age dominated the regression. Self-reports are partial on both sides. Two asymmetries nonetheless remain. At least one human cue carried genuine evidential value, whereas model rationales tracked the answer rather than the outcome; and humans freely admitted uncertainty (49% endorsing feeling/intuition), whereas expressions of uncertainty were essentially absent from model rationales.

The comparison also reveals what transfers between humans and machines: difficulty structure. Fifteen of 19 models found ChatGPT-4o portraits harder than Imagen 3 portraits, as humans did (one, claude-fable-5, tied the two at 100%). This suggests that within this stimulus pool, detectability is chiefly a property of the generator output itself. Model choice, by contrast, mattered enormously and unpredictably. A dedicated vision–language architecture outperformed its newer general-purpose sibling by 13.6 pp, and a mid-tier model outperformed the flagship of its own family by 22.7 pp. Benchmark reputation transferred poorly too: kimi-k3, reported at the top of contemporary coding leaderboards, placed 12th of 19 here. Capability rankings from one domain evidently do not predict perceptual forensics. In this roster, neither price, recency, nor fame on other benchmarks guaranteed detection ability.

Several limitations bound these conclusions. The benchmark inherits the scope of the human study: one portrait domain, two generator families frozen in July 2025, and a single-image REAL/AI task with balanced base rates. Results are a time-stamped snapshot, not a stable ranking of generators or models [13]. We tested a single prompt family (one instruction, varied examples) at temperature 0 through one aggregator service (OpenRouter). Different instructions, decoding settings, or provider-side routing could shift absolute numbers, though our robustness analysis suggests the example-sensitivity we quantify would remain. Running a model at temperature 0 on a hosted service also does not guarantee identical output every time; the serving system adds some randomness. Our identical-prompt control — run 15 days later, so it combines serving nondeterminism with any provider-side drift — bounds this noise at 1.5–2.0% of answers for the most stable models and 9.1–11.6% for the OpenAI models. In matched two-pass comparisons the example effect clearly exceeds the noise for the stable models but only barely for the OpenAI models; the control covers four June-cohort models, so this decomposition extends neither to the rest of that cohort nor to the July additions. Model training data are opaque. FFHQ is a public dataset, so some real photographs (and conceivably descriptions of them) may appear in model training corpora. We cannot rule out contamination, which would if anything inflate model performance and strengthen our headline conclusion. The stimulus files themselves (including the AI counterparts) have moreover been publicly served by that study's web application since August 2025, and its public data release exposes each stimulus's class through its file structure. Labelled stimulus–class pairs could therefore appear in post-2025 training corpora. This too would be expected to inflate, not deflate, model performance. Response latency of API calls reflects computation and serving infrastructure, not perceptual processing time, so no analogue of the human reaction-time analyses is possible (Supplementary Fig. S4). Finally, our persona-prompting screen was deliberately small. It licenses only the narrow conclusion that a minimal age-persona manipulation does not reproduce the human age gradient in these models.

These caveats do not soften the central pattern. Under a protocol designed for exact comparability, the frontier crossed the young-adult human ceiling between June and July 2026: sensitivity that no human age group matches, from two independent model families, surviving averaging over prompt draws. What the leading machines have not yet borrowed from people is the quiet part of perception — a criterion held at neutral, judgments stable under irrelevant context, and reasons that are evidence rather than echo. The question "can AI detect AI?" now has a new answer — better than the average adult at any age we measured — and a new caveat: the systems that see best still do not decide like one.

# Material and methods

## Human reference data

No new human data were collected in this study. The human reference data come from an earlier study by two of us [13], reported in a preprint that has not been peer reviewed (arXiv:2603.24048) and approved by the Institutional Review Board of Hwasung Medi-Science University (HSMUIRB-2025-06). Every human value reported here — accuracy, class and generator accuracy, confidence association, and the signal-detection coordinates — is our own recomputation from that study's public de-identified data release, not a figure taken from the preprint; the preprint's reported values are given alongside ours in table 3 for comparison. We describe the design in enough detail for the present benchmark to stand on its own.

Participants judged portraits in a web-based task, at their own pace, between August 2025 and February 2026. Recruitment was public and used snowball sharing through online communities and social media; participation was voluntary under digital informed consent, with no monetary or material compensation (participants received only a score report and a non-monetary "title" image after finishing). The interface was available in Korean and English; instructions and survey items followed the participant's language choice, which was made after the practice block (practice screens were shown bilingually), while the response buttons were fixed as "REAL"/"AI", and stimuli and task structure did not vary by language. The sample is therefore self-selected and strongly mobile-skewed, and viewing conditions (screen, distance, lighting) were uncontrolled.

Each trial began with a central fixation cross (1,000 ms), then one portrait; participants chose REAL or AI with two on-screen buttons, untimed. This is a single-interval binary (yes/no) judgment, not a two-alternative forced choice between simultaneously presented images, so sensitivity is estimated with the yes/no d′ formula (Statistics). After 4 practice trials with feedback (2 real, 1 ChatGPT-4o, 1 Imagen 3), each participant judged 20 portraits with no feedback. The 20 main stimuli were sampled at random per session with a fixed composition (10 real, 5 ChatGPT-4o, 5 Imagen 3); identities used in that session's practice were excluded, and no identity was repeated within the main trials. Images were displayed at up to 450 px (350 px on narrow screens) with zooming disabled. Our analytic sample applies the two filters of that study to its public release: keep each person's self-reported first attempt, and keep ages 20–69. Of 1,843 logged records this leaves N = 1,667 (mobile n = 1,332; PC n = 335). No further exclusions were applied; repeat participation was identified by self-report only, not by device or network fingerprinting. The preprint's own run of the same filters gave N = 1,664, a three-participant difference that its repository attributes to an independently reconstructed de-identified export while recording the exact cause as unresolved. Overall accuracy differs by 0.06 pp (85.18% vs 85.24%). Mean accuracy was 85.18% (median 90%). Accuracy by age group, with per-group sample sizes and standard deviations, was 88.5% (20s; n = 764, SD 10.0), 87.7% (30s; n = 534, SD 10.2), 80.5% (40s; n = 169, SD 14.2), 71.8% (50s; n = 121, SD 17.3) and 65.9% (60s; n = 79, SD 16.7). Real-photo and AI-detection accuracy were nearly equal (85.7% vs 84.7%); by generator, accuracy was 80.4% on ChatGPT-4o images and 88.9% on Imagen 3 images. After the main trials, participants reported their confidence (5-point scale). They also ticked, on a fixed multiple-response checklist, the visual cues they had used: hands/fingers, eyes and other facial detail, background, skin or fabric texture, painting-like or artificial quality, unrealistically perfect lighting, idealized appearance, unnatural symmetry, abnormal text, an overall unsettling feeling, plus "don't know", "random guess" and a free-text option. These give the confidence–accuracy association (Spearman $\rho$ = 0.379, n = 1,667) and the strategy endorsement rates in Fig. 6c. One caveat applies to those rates: 202 participants (12.1%) have every checklist item except the free-text option marked, together with a stray `on` token from the form's select-all control — a serialization fault rather than genuine endorsement. Excluding them lowers the rates of the twelve affected items without changing their ordering (e.g., painting-like 68.6% → 64.2%, texture 65.3% → 60.5%, eyes 42.8% → 34.9%; the free-text option, which the fault does not touch, rises from 1.2% to 1.4%), so we report the unfiltered rates and treat all human–model cue comparisons as qualitative. For the head-to-head comparison we also place the human observers on the same signal-detection axes, recomputed from the public data (Statistics). They were essentially unbiased at every age (criterion $c \approx 0$), while sensitivity $d'$ fell steeply with age (participant-mean $d'$ = 2.33, 2.27, 1.76, 1.24, 0.87 for the 20s through 60s). These human signal-detection coordinates serve only to position humans against the models; we do not attempt an in-depth account of human factors here. Age-group 95% CIs were computed as ±1.96 standard errors of the mean (SEM) within each age bin. The human cue analysis quoted in the Results is also ours: a binomial logistic regression of each participant's 20-trial score on the 13 checklist dummies with age (centred), sex and device as covariates, with HC3 (leverage-adjusted) heteroscedasticity-robust standard errors, reported as odds ratios and as average marginal effects. It is fitted to the 1,465 participants whose checklist responses are free of the select-all fault described above. Absorbing the 202 faulty rows with an indicator instead gives very similar coefficients and the same qualitative inference (texture OR 1.35), whereas entering them untreated makes the painting-like coefficient nominally significant ($p$ = .04) — which is why the faulty rows are excluded rather than absorbed. The robust covariance is computed directly (leverage-adjusted HC3 sandwich) rather than delegated to a library, because a widely used implementation silently returns HC0 when HC3 is requested; the deposited script asserts that the leverage adjustment is present. The script is deposited with the analysis code and runs on the human study's public CSVs.

## Stimuli

The stimulus pool was identical to that of ref. [13]: 210 face portraits (512×512 px), comprising 70 real photographs sampled from FFHQ [1] and 140 AI counterparts. The AI images were generated from the same 70 identities in July 2025: 70 by ChatGPT-4o native image generation [38] and 70 by Imagen 3 via Gemini 2.5 [39]. We used a mirroring (reverse-prompting) procedure. Each generator first wrote its own prompt describing the source portrait, then used that prompt to recreate it. As in the human experiment, a small white patch occludes the bottom-right corner of every image (real and AI) to mask trivial corner artefacts. Four base identities served as labelled practice/few-shot images (2 real, 1 ChatGPT-4o, 1 Imagen 3). All three versions of each practice identity were excluded from the main trials, leaving 198 main stimuli (66 identities × 3 sources) and guaranteeing that no identity shown as a labelled example appeared among the judged items. Images were transmitted to models as base64-encoded data URLs at the deployed 512×512 resolution; filenames, metadata, and any information about source or answer were never sent.

## Models

Nineteen vision-capable models from nine providers were evaluated in two cohorts (Table 1). The June-2026 cohort comprised 14 models: gpt-5.5, gpt-5.4, gpt-5.4-mini (OpenAI); claude-opus-4.8, claude-sonnet-4.6 (Anthropic); gemini-3.1-pro-preview, gemini-3.5-flash, gemma-4-31b-it (Google); grok-4.3 (xAI); qwen3-vl-235b-a22b-instruct, qwen3.5-flash-02-23 (Alibaba); glm-4.6v (Zhipu); mistral-medium-3.1 (Mistral); and llama-4-maverick (Meta). The July-2026 cohort added five releases that appeared after the first collection: gpt-5.6-sol, gpt-5.6-terra, gpt-5.6-luna (OpenAI); kimi-k3 (Moonshot); and claude-fable-5 (Anthropic). All models were queried through the OpenRouter application programming interface (API) with the exact model identifiers listed in Table 1 — the June cohort on 22–23 June 2026 (single-pass and iteration runs) and 7 July 2026 (persona pilot), the July cohort on 20 July 2026 (single-pass, iteration, and effort-pilot runs) — at temperature 0 with a 2,000-token output cap (raised to 8,000 in the reasoning-effort pilot) and a 120-s request timeout. Provider-side routing of open-weight models is determined by OpenRouter and was not constrained. Models are referred to in the text by shortened display names; Table 1 gives the exact identifiers.

## Benchmark protocol

Each request comprised (i) a fixed system prompt ("You are a careful image forensics evaluator participating in an AI-image-detection benchmark. You judge only the visible image content."), (ii) the four labelled practice images presented as few-shot examples with their correct labels — mirroring the human practice phase, in which feedback was given — and (iii) one target image with a fixed user instruction, quoted verbatim: 'Decide whether this image is a REAL photograph or an AI-GENERATED image. Respond with ONLY compact JSON and nothing else: {"answer":"REAL" or "AI","confidence":<integer 0-100>,"rationale":"<one short sentence>"}.' The complete request structure (message order, few-shot formatting, and API parameters) is provided in the electronic supplementary material. Each model judged each of the 198 main stimuli exactly once per pass. Calls were independent (no conversational memory across trials), matching the absence of feedback during human main trials. Temperature 0 makes each model pick its most likely output, so answers vary as little as possible. Even so, identical output is not fully guaranteed, because the models run on outside servers we do not control. Responses answering neither REAL nor AI were marked unclear (23 of 3,762 responses in the main benchmark; 21 of these came from kimi-k3 hitting the output-token cap) and excluded from accuracy denominators. The main single-pass benchmark comprised 198 × 19 = 3,762 judgments with zero API errors. Of the 3,739 valid responses, 3,738 included a non-empty rationale (one omitted it); these constitute the rationale corpus analysed below.

## Human-comparable scoring

Because human sessions contained 10 real, 5 ChatGPT-4o and 5 Imagen 3 trials, model accuracy is reported as balanced accuracy = 0.5·(real accuracy) + 0.25·(ChatGPT-4o accuracy) + 0.25·(Imagen 3 accuracy), the expected per-session accuracy of an observer with the model's class-specific accuracies. For distribution-level comparisons, we bootstrapped 20-trial sessions (10/5/5 composition; 10,000 resamples; fixed seed) from each model's per-item outcomes. Class-specific 95% CIs used the Wilson score interval [41]; the CI for balanced accuracy propagated the three binomial variances with the session weights. The three versions of one identity (its real, ChatGPT-4o and Imagen 3 images) are related, not independent. So we re-checked every headline confidence interval with a cluster bootstrap — resampling the 66 base identities together with their three versions (10,000 resamples). These clustered intervals were within ~1 pp of the item-level ones (half-widths 3.0–5.9 pp across the 19 models; e.g. gpt-5.5 [83.0, 91.3]; gpt-5.6-sol [89.8, 95.8]). The resampling stream is fixed and defined over the released data alone: identities are iterated in the sorted order of their frozen released codes and resampled with Python's Mersenne Twister (`random.Random(2026).choice`), re-seeded for each model, with interval endpoints taken as the 250th and 9,750th order statistics of the 10,000 resamples. The released package therefore reproduces every quoted endpoint exactly (Data accessibility); a different generator satisfying the same description would move endpoints by up to ~0.4 pp. Pairwise model differences were tested with a two-sided exact binomial McNemar test (nominal, unadjusted) on the shared valid items (176–198 per pair after excluding unclear responses; the low end reflects kimi-k3's 21 unclears) [42]. The exact form matters at the margin: one non-headline pair is separated by the exact test ($p = .049$) but not by the continuity-corrected $\chi^2$ approximation ($p = .050$). Non-significance is reported descriptively as a power-limited separability screen, not as evidence of equality; multiplicity correction across the 171 pairwise tests would only enlarge the reported indistinguishable set (the reported separations of the July leaders from the June leaders are nominal, unadjusted values). Response bias is summarized as AI-detection accuracy minus real-photo accuracy; the said-AI rate is the fraction of valid parsed answers equal to "AI". Overconfidence is mean stated confidence minus balanced accuracy; resolution is mean confidence on correct minus incorrect answers; calibration is additionally summarized by the Brier score [43] of stated confidence against correctness. Mean confidence is an unweighted average over each model's valid responses carrying a numeric rating (n = 177–198; 198 for 16 models, 197 for the two Gemini models and 177 for kimi-k3); recomputing it with the session weights changes the overconfidence values by at most 1.3 pp.

## Robustness to few-shot examples (iteration design)

To measure sensitivity to the labelled examples — the only element of the protocol we vary — we repeated the full benchmark five times per model with varied practice draws. For iteration k (seed-derived, deterministic), 4 of the 70 base identities were drawn as practice (2 real, 1 ChatGPT-4o, 1 Imagen 3), their triplets excluded, and the remaining 198 stimuli judged once each, exactly as in the main benchmark (n = 198 per draw; 18,810 additional judgments in total across the 19 models; zero errors). The main benchmark's hand-selected practice set thus constitutes a sixth, baseline draw with identical design. We report the mean ± SD and range of balanced accuracy over the five varied draws (the baseline draw is shown separately, as it was not randomly selected). We also report the per-item flip rate: the fraction of items whose answer changed at least once across the six passes in which they appeared (the baseline plus the five varied draws). Counting only strict REAL↔AI reversals gives 24.7% fleet-wide, and strict reversals across the five varied draws alone give 23.3%, versus the headline any-change 25.7% (24.2% for any-change across the five varied draws). To quantify run-to-run noise directly, we reran the unchanged baseline prompt on all 198 items for four June-cohort models spanning the stability range (gpt-5.4, gpt-5.5, claude-sonnet-4.6, qwen3-vl-235b; 792 calls, zero errors). This control ran 15 days after the original collection, so it combines serving nondeterminism with any provider-side drift over that interval. Answers changed for 1.5% (claude-sonnet-4.6), 2.0% (qwen3-vl-235b), 9.1% (gpt-5.5) and 11.6% (gpt-5.4) of items. For a like-for-like comparison with the example manipulation we use the same two-pass statistic: the mean disagreement between the baseline draw and each varied draw was 6.8% (claude-sonnet-4.6) and 6.9% (qwen3-vl-235b) — 3.4–4.5× their noise — but 11.4% (gpt-5.5) and 14.4% (gpt-5.4), only 1.2–1.3× their noise. The example effect is therefore clearly real for the stable models, whereas most of the OpenAI models' measured instability is intrinsic run-to-run noise. The six-pass flip rates in the Results include both sources, and this decomposition is restricted to the four models tested.

The SD over draws is an upper bound on example-sensitivity of accuracy, as the judged 198-item set also shifts slightly between draws (4 held-out identities differ). Computing flips only on identical items removes the change in test-set composition; the rate still combines example sensitivity with run-to-run noise.

## Rationale cue coding and diagnosticity

The 3,738 rationales from the main benchmark were coded into a 15-category cue taxonomy (over-smoothness, skin texture, lighting/shadow, eyes, teeth/mouth, hair, ears, background, symmetry/proportion, artifacts/distortion, colour/tone uniformity, accessories, natural/realistic impression, uncanny/too-perfect, intuition/uncertainty) by an LLM coder (gpt-5.4-mini, temperature 0). Coding was multi-label and blind to the true class and to whether the judgment was correct. The taxonomy was constructed to cover both the cue categories of the human study's strategy checklist [13] and cues salient in model rationales. The coder is itself one of the benchmarked models. We mitigated the resulting circularity by blinding it to ground truth and correctness, and by cross-checking against a rule-based keyword coder, which produced consistent per-model cue profiles (Supplementary Fig. S5). Cue frequency is the percentage of rationales mentioning a cue. Cue diagnosticity is the accuracy of judgments whose rationale invokes the cue, minus the base accuracy of all judgments (Δ base), computed overall and within true class (Δ within AI; Δ within real). Verdict-conditioned cue validity was computed as the accuracy difference, among judgments sharing the same verdict, between those whose rationale cites a cue and those whose rationale does not. Cell sizes for every cue × class and cue × verdict cell — for both the cue-citing and non-citing groups — are given in the electronic supplementary material. The class-conditioned deltas quoted in the text rest on cells of $n \geq 138$; the verdict-conditioned natural/realistic comparison rests on a small comparator cell (62 REAL verdicts without the cue) and is quoted with its uncertainty.

## Reasoning-effort pilot

To ask whether deeper deliberation aids perceptual discrimination, we reran the single-pass benchmark for the July-cohort leader gpt-5.6-sol with its reasoning effort set to the maximum ("xhigh") via the API's reasoning parameter, raising the output cap to 8,000 tokens to accommodate reasoning traces (198 calls, zero errors, 20 July 2026); everything else was unchanged. Relative to its default run, the xhigh condition changed 6.6% of answers, raised balanced accuracy from 92.8% to 94.3%, and eliminated false alarms entirely (real-photo accuracy 100%; ChatGPT-4o detection 81.8%, Imagen 3 95.5%; mean confidence 92.7). The gain was modest and even: each of the three source classes improved by exactly one image, and on AI trials the answer changes were seven corrections against five regressions. As a single-model, single-condition pilot this is reported descriptively.

## Age-persona prompt-sensitivity pilot

A small screening pilot tested whether demographic role-play shifts model behaviour (Supplementary Fig. S3). Four models spanning the accuracy/bias space (qwen3-vl-235b, gpt-5.5, claude-sonnet-4.6, llama-4-maverick) judged a fixed 60-stimulus subset (20 identities × 3 sources) under two persona system prompts that replace the neutral evaluator framing ("You are answering as a typical human participant in their 20s [60s]. You are participating in an AI-image-detection benchmark. Judge only the visible image content. Do not use file names, metadata, external knowledge, or assumptions about the experiment. Return a careful forced-choice judgment."), with the user prompt, few-shot examples, and output format unchanged. The two persona conditions required 480 new API calls (zero errors; run 7 July 2026). The neutral comparator reused the main benchmark's judgments on the same 60 stimuli. Persona-versus-neutral contrasts therefore reflect the whole system-prompt change, whereas the 20s-versus-60s contrast isolates the age clause. Pre-specified expansion criteria (≥5 pp balanced-accuracy shift, ≥10 pp said-AI shift, or ≥10 pp confidence shift for any model) were not met (maxima: 3.8 pp, 3.3 pp, 3.8 pp; per-item flips ≤6.7%), so the pilot was not expanded. The four models constitute a cost-aware screening sample chosen for informative contrast, not a representative sweep. The manipulation tests prompt sensitivity, not simulation of human ageing.

## Statistics

All tests are two-sided with $\alpha = 0.05$. Confidence intervals for proportions use the Wilson score interval [41]; paired model comparisons use the two-sided exact binomial McNemar test [42]; session-level distributions use the nonparametric bootstrap [44] (10,000 resamples, fixed seed); calibration uses the Brier score [43]. Our signal-detection analysis treats "AI" as the signal to detect. The hit rate is AI-detection accuracy. The false-alarm rate is one minus real-photo accuracy. Sensitivity is $d' = z(\text{hit}) - z(\text{FA})$, and response criterion is $c = -\tfrac{1}{2}\cdot[z(\text{hit}) + z(\text{FA})]$, where $z$ is the inverse normal; a negative $c$ means a liberal, AI-leaning bias. For humans, hit and false-alarm rates were computed for each participant from their 10-real/10-AI session, using the log-linear correction (adding 0.5 to each count so that perfect scores do not break the formula), then averaged within age group. Model $d'$ and $c$ are reported with the same log-linear correction applied to each model's pooled single-pass counts (66 real; 111–132 valid AI trials — kimi-k3's 21 unclear responses all fell on AI trials). Human pooled-count $d'$ applies the correction to the summed counts within each age group (20s: 2.41; 30s: 2.32). For models away from ceiling the correction is small (≤0.07 on $d'$; e.g. gpt-5.5: 2.49 → 2.42), but it is essential at ceiling: claude-fable-5's perfect hit rate makes the uncorrected $d'$ (5.50) and criterion (−2.00) clamp artefacts, versus corrected values of 3.41 and −0.97. Model $d'$ uncertainty was assessed with the identity-level cluster bootstrap using the corrected estimator and the same fixed resampling stream defined over the released identity codes (percentile convention: the 250th/9,750th order statistics of 10,000 resamples; endpoints vary by ~±0.05 across conventions and resampling streams). Resulting 95% CIs: gpt-5.6-sol [2.65, 3.75], claude-fable-5 [3.10, 3.75], June leaders [2.0, 3.2]. The SD reported over iteration draws is the population SD of the five draw means. Analyses were performed in Node.js and Python 3 with standard scientific libraries; all analysis code is available (see Data accessibility).

## Ethics

The human reference data were reused from the earlier study of ref. [13], which was approved by the Institutional Review Board of Hwasung Medi-Science University (approval no. HSMUIRB-2025-06); participants provided digital informed consent and no identifying information was collected. No new human-subjects data were collected for the present work, which evaluated publicly available AI models through a commercial application-programming interface and required no additional ethical approval.

## Data accessibility

Aggregate results — per-model, per-class and per-condition summaries, the signal-detection values and the cue cell sizes (tables 1–6) — are provided as electronic supplementary material and deposited at https://github.com/gdrpaul3-byte/hsmu_vlm_detection_benchmark (release v1.0.0, archived at https://doi.org/10.5281/zenodo.22148304) [45]. Every value plotted in the figures is either in these tables or directly reconstructible from the released per-item files (e.g., the full pairwise McNemar matrix of fig. S2, the persona condition summaries of fig. S3 and the per-model cue profiles of fig. S5). The stimulus pool, its FFHQ image-level attribution and licence metadata (supplementary table S1 of ref. [13]), and the de-identified human data are available in that study's public repository (https://github.com/gdrpaul3-byte/hsmu_ai_detection_public; release v1.0.1, the version analysed here). Collection, iteration, rationale-coding, statistics and figure-generation code, a data dictionary and environment details are deposited alongside the data. A standalone script reproduces the headline statistics directly from the released per-item files and self-checks against the master table; re-rendering the figure images additionally requires the stimulus images from that study's repository and the local raw runs. Per-item model judgments are deposited in full — answer, confidence and correctness for the main benchmark, the iteration draws, the identical-prompt control, the persona pilot and the reasoning-effort pilot, plus the 15-category cue codes of each main-benchmark rationale and anonymized identity-cluster codes — under randomized stimulus codes that cannot be joined to the public stimulus files. Subject to the editors' agreement, two categories of material are withheld. The stimulus-code-to-filename mapping is withheld so that the deposited tables do not become a machine-readable answer table for the still-deployed benchmark application (each stimulus's class is recoverable from that study's public release by manually matching images, but not in-band during a test). The collection seeds and production stimulus identifiers are withheld with it — and replaced by placeholders in the released code copies — because they could partially regenerate the mapping. The verbatim rationale texts are withheld because they describe image content; their full cue codes are released. Both are available from the corresponding author. Publication figures necessarily reveal the ground truth of the few example stimuli they display (figures 1a and 6a; as does fig. 1 of ref. [13]). For the three judgments displayed in fig. 6a, the shown verdicts, confidences and cues also allow the corresponding released item codes — and, through the identity codes, their six counterparts — to be identified; we accept this bounded nine-item disclosure as the cost of showing worked examples.

## Declaration of AI use

Generative AI tools were used in preparing this manuscript. The authors used a large language model (Anthropic Claude) to assist with drafting and editing prose, writing analysis and figure-generation code, and formatting; all study design, analyses, numerical results and interpretations were checked and verified by the authors, who take full responsibility for the content. Language models were also used as adversarial reviewers during verification: over ten review rounds, multi-agent panels (Anthropic Claude) and an independent agent based on OpenAI Codex (GPT-5.6) re-derived the reported statistics from the raw and deposited data, probed the public release for re-identification risk, and audited the manuscript for internal consistency. These tools raised candidate issues only; each issue and each number was re-checked by the authors against their own analysis pipeline before any revision was made. Separately, and as documented in the Material and methods, a large language model (gpt-5.4-mini) was used as a described analysis component to code the free-text model rationales; this is part of the reported methodology rather than manuscript preparation. For transparency: one evaluated model (claude-fable-5) belongs to the same model family as the writing-assistance tool. The evaluation itself was conducted through the OpenRouter API under the identical protocol applied to all models, with no connection to any authoring session.

## Authors' contributions

S.W.K.: conceptualization, methodology, software, data collection, formal analysis, writing—original draft; S.Y.K.: methodology, formal analysis, writing—review and editing; M.J.: methodology (design of selected experiments), interpretation, writing—review and editing; J.T.: methodology (design of selected experiments), interpretation, writing—review and editing. All authors gave final approval for publication and agree to be held accountable for the work performed therein.

## Conflict of interest declaration

We declare we have no competing interests.

## Funding

This work was supported by the HSMU Research Grant (2025).

## Supplementary Information

Supplementary figures accompany the main text; each appears on its own page below. All human values are recomputed from the public data release of ref. [13]; model results are aggregate summaries.

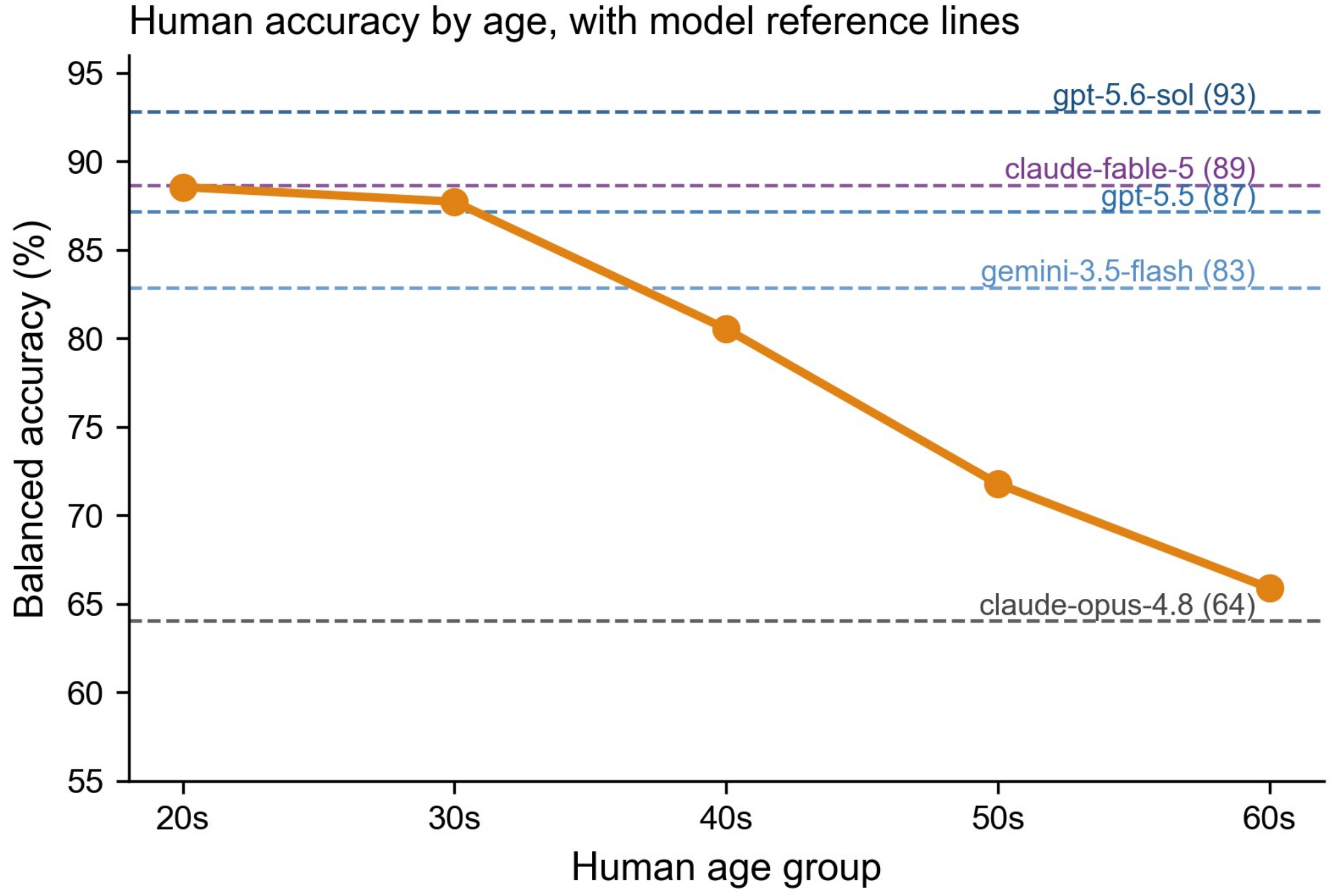

**Supplementary Figure S1 | Human accuracy by age.** Human accuracy by age group with model reference lines; the July leaders (gpt-5.6-sol, claude-fable-5) sit above the 20s band; June leaders near it. Human data reused from ref. [13].

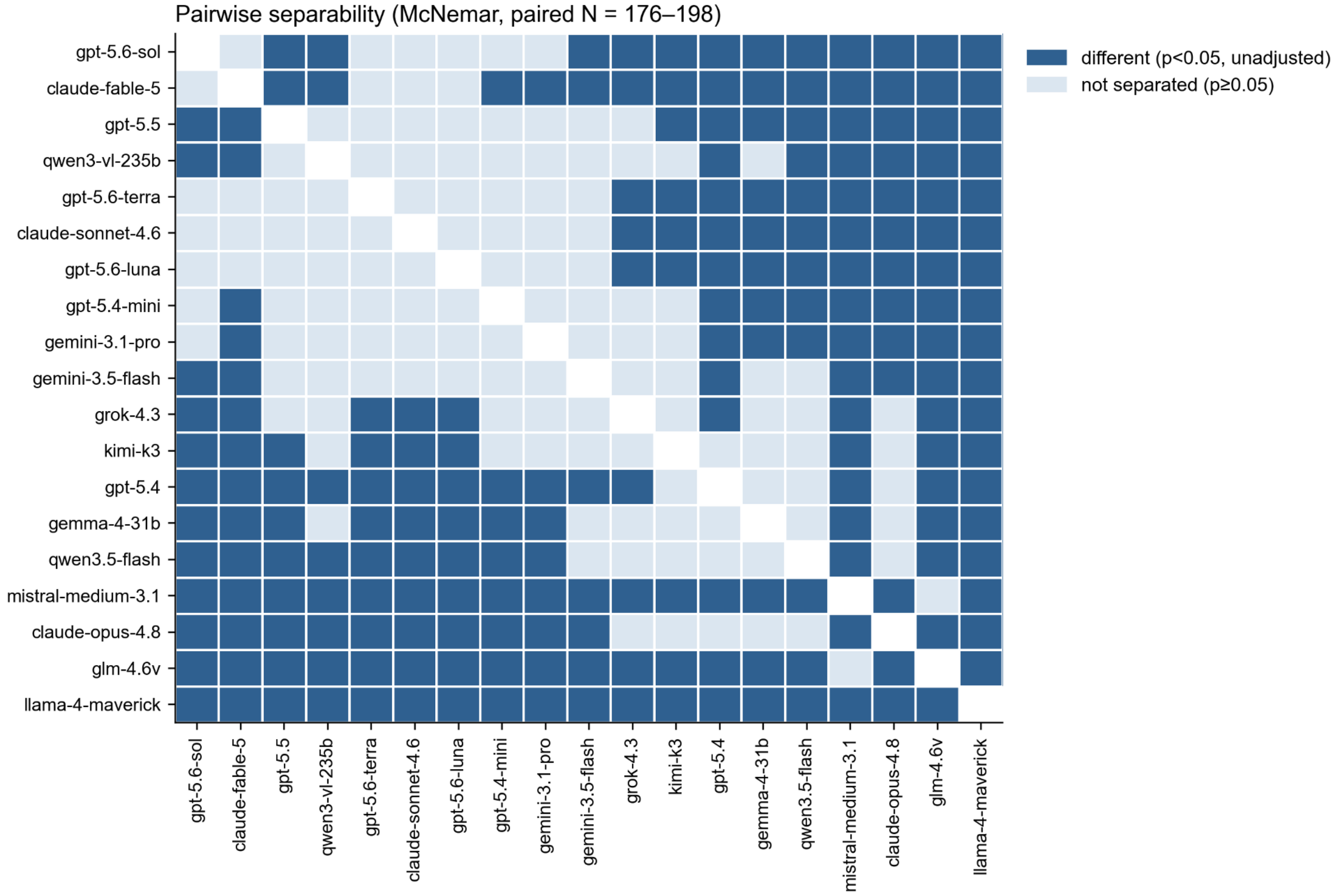

**Supplementary Figure S2 | Pairwise McNemar separability.** Pairwise McNemar tests on the shared valid items (176–198 per pair after excluding unclear responses; unadjusted). The July leaders separate from the June leaders (gpt-5.6-sol and claude-fable-5 each beat gpt-5.5 and qwen3-vl-235b, $p < .05$) but not from each other.

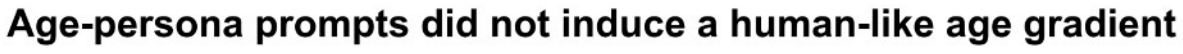

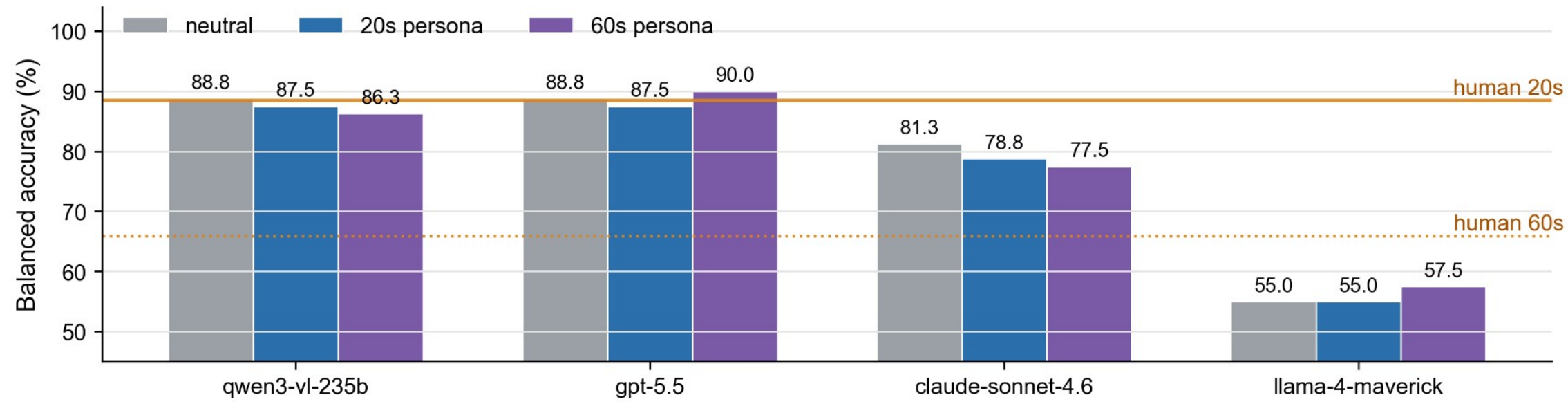

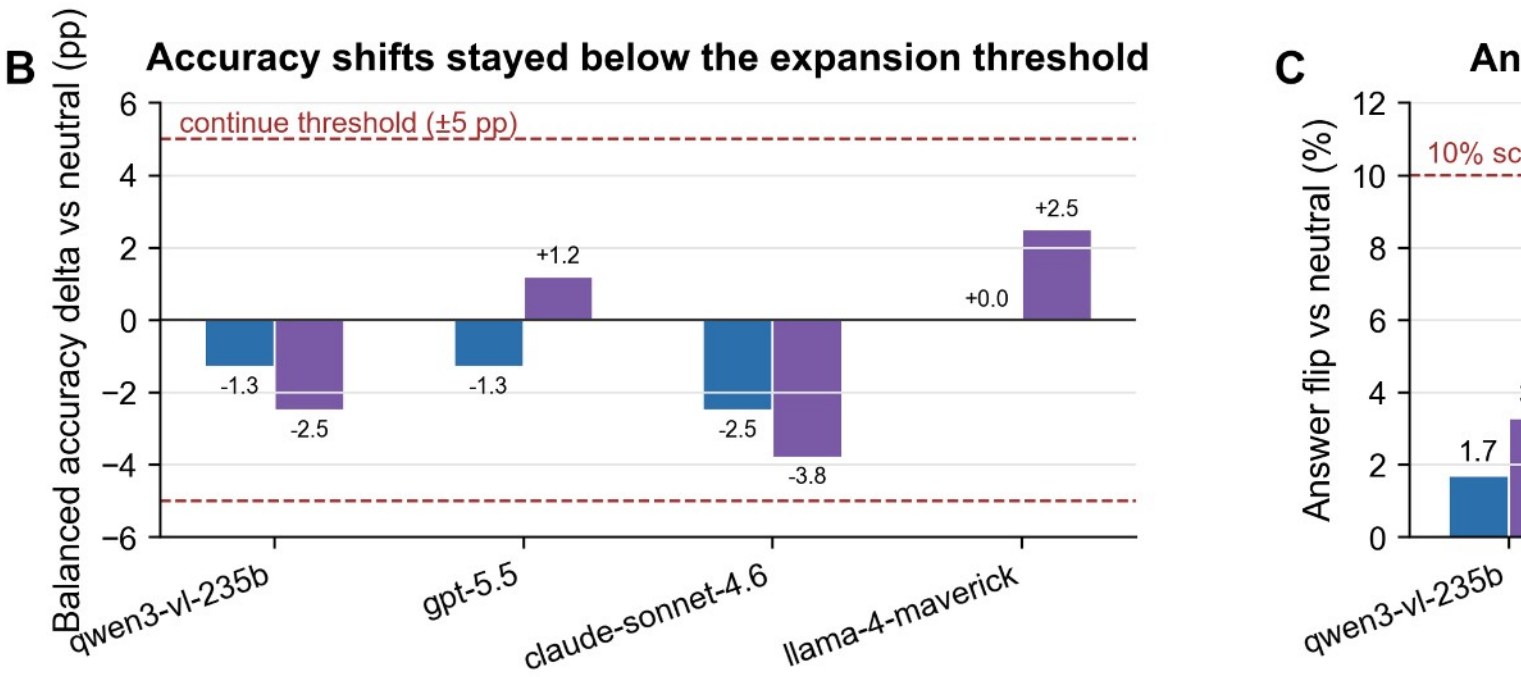

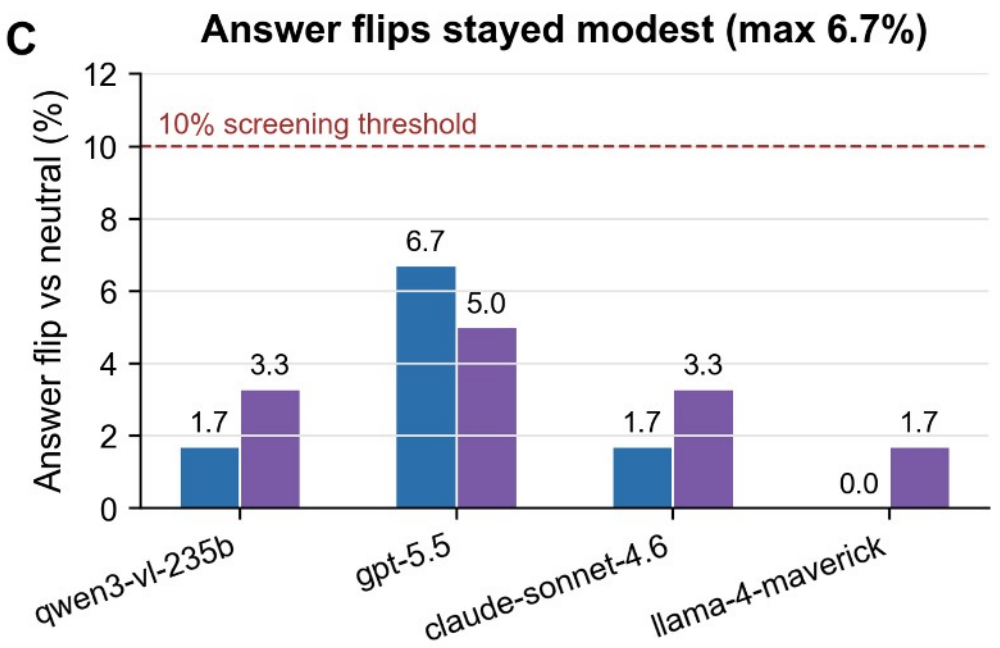

**Supplementary Figure S3 | Age-persona prompt-sensitivity pilot.** Four models × two personas (20s, 60s) × 60 stimuli. Balanced-accuracy shifts ≤3.8 pp, answer flips ≤6.7%, and no human-like age gradient; the pilot did not meet pre-specified expansion criteria. Human 20s/60s reference lines shown.

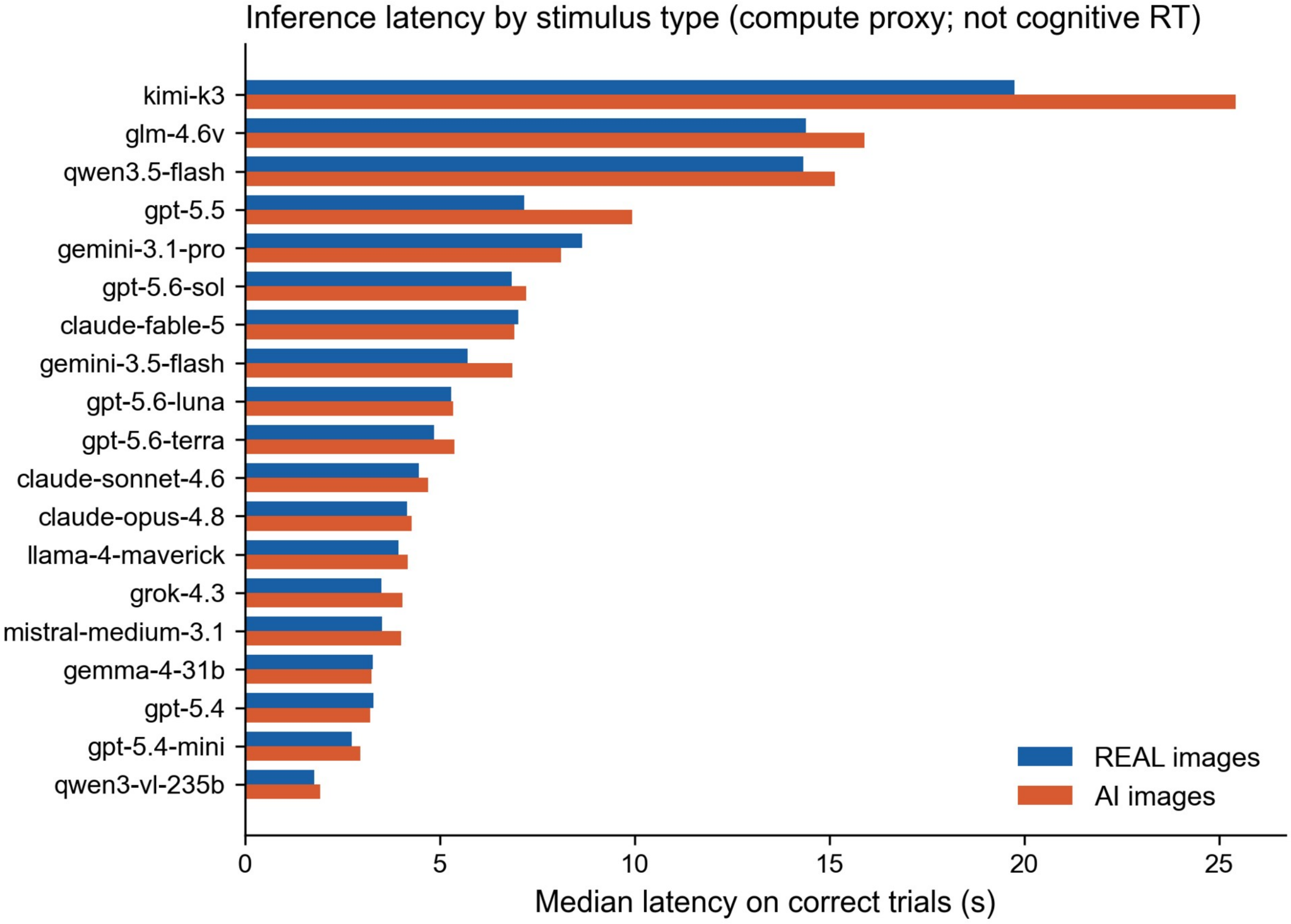

**Supplementary Figure S4 | Inference latency by stimulus type.** Model API response latency by stimulus class. Latency is an API/compute proxy and is not comparable to human reaction time; it is reported only for completeness.

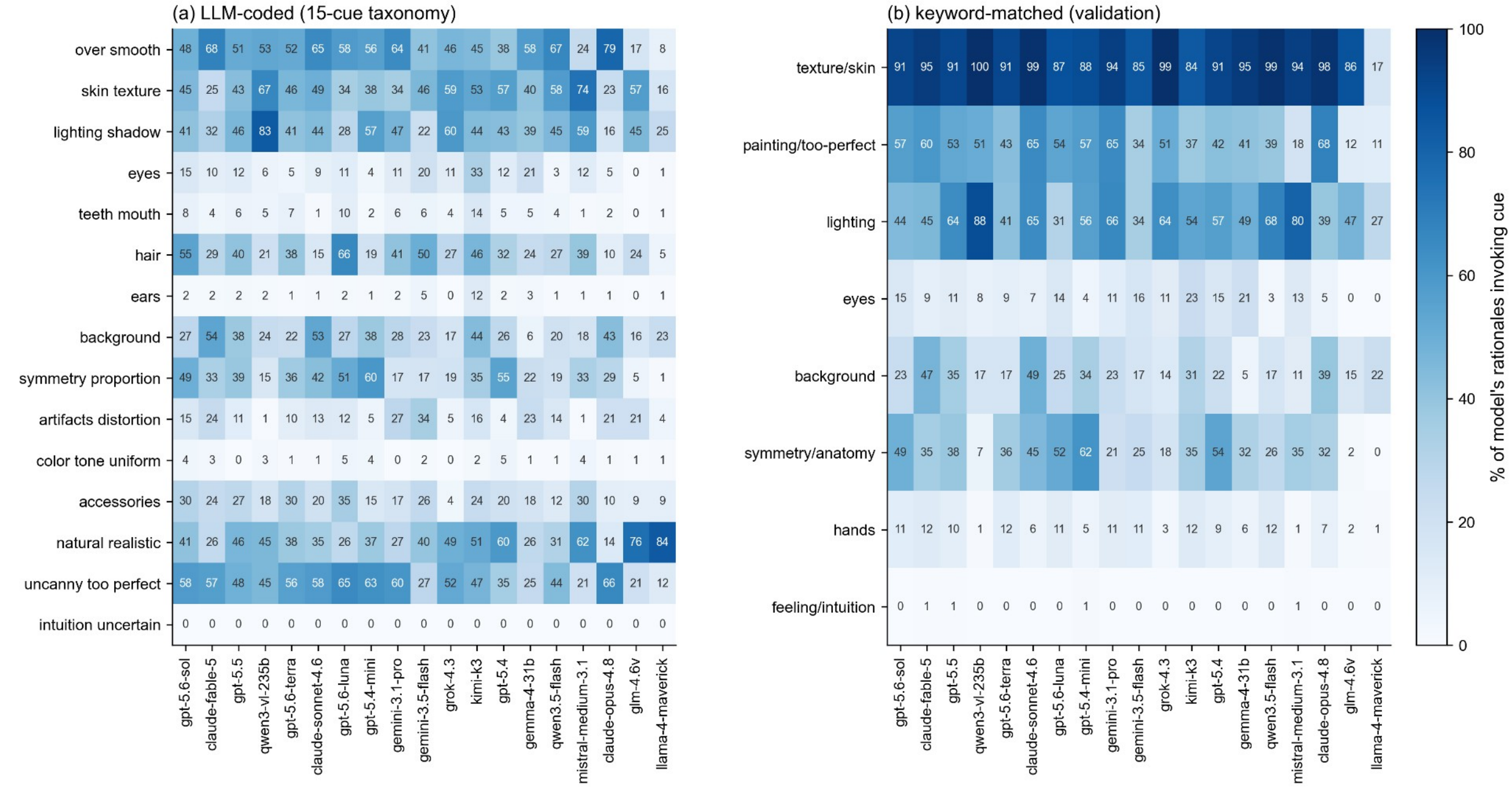

**Supplementary Figure S5 | Cue fingerprints per model.** Per-model cue frequencies under (a) LLM-based coding and (b) rule-based keyword matching; the two coding methods yield consistent per-model profiles (method robustness).